\documentclass[referee,a4paper,12pt,traditabstract]{swsc} 

\usepackage{graphicx}
\usepackage{txfonts}
\usepackage{subfigure}
\usepackage{epstopdf}
\usepackage[mathlines]{lineno}
\usepackage[authoryear,round]{natbib}
\usepackage[backref]{hyperref}
\usepackage{url}
\usepackage{lscape}
 
\hypersetup{colorlinks=true,citecolor=cyan,urlcolor=cyan,linkcolor=blue}

\begin{document}


\title{Is the $\mathrm F_{10.7cm}$\,--\, Sunspot Number relation linear and stable?}

   
\titlerunning{$\mathrm F_{10.7cm}$ proxy relation and temporal homogeneity}
\authorrunning{Clette}

\author{F. Clette}
\institute{World Data Center SILSO, Royal Observatory of Belgium, 1180 Brussels,   Belgium\\
\email{\href{mailto:frederic.clette@oma.be}{frederic.clette@oma.be}}}

 
\abstract
{The $F_{10.7cm}$ radio flux and the Sunspot Number are the most widely used long-term indices of solar activity. They are strongly correlated, which led to the publication of many proxy relations allowing to convert one index onto the other. However, those existing proxies show significant disagreements, in particular at low solar activity. Moreover, a temporal drift was recently found in the relative scale of those two solar indices. 

Our aim is to bring a global clarification of those many issues. We compute new polynomial regressions up to degree 4, in order to obtain a more accurate proxy over the whole range of solar activity. We also study the role of temporal averaging on the regression, and we investigate the issue of the all-quiet $F_{10.7}$ background flux. Finally, we check for any change in the $F_{10.7}$\,--\,sunspot number relation over the entire period 1947--2015.

We find that, with a $4^{th}$-degree polynomial, we obtain a more accurate proxy relation than all previous published ones, and we derive a formula giving standard errors. The relation is different for daily, monthly and yearly mean values, and it proves to be fully linear for raw non-averaged daily data. By a simple two-component model for daily values, we show how temporal averaging leads to non-linear proxy relations. We also show that the quiet--Sun $F_{10.7}$ background is not absolute and actually depends on the duration of the spotless periods. Finally, we find that the $F_{10.7cm}$ time series is inhomogeneous, with an abrupt 10.5\% upward jump occurring between 1980 and 1981, and splitting the series in two stable intervals. 

Our new proxy relations bring a strong improvement and show the importance of  temporal scale for choosing the appropriate proxy and the $F_{10.7}$ quiet-Sun background level. From historical evidence, we conclude that the 1981 jump is most likely due to a unique change in the $F_{10.7}$ scientific team and the data processing, and that the newly re-calibrated sunspot number (version 2) will probably provide the only possible reference to correct this inhomogeneity.}


   \maketitle

\section{Introduction}
The sunspot number (hereafter SN; symbol: $S_{\mathrm{N}}$ ) \citep{{CletteEtal2014},{CletteLefevre2016}} and the $\mathrm F_{10.7cm}$ radio flux (symbol: $F_{10.7}$) \citep{TappingMorton2013} are arguably the most widely used solar indices to characterize the long term evolution of the solar activity cycle and of the underlying dynamo mechanism. In order to be usable over duration of decades to centuries, those indices must guarantee a long term stability so that levels of activity at two widely spaced epochs can be compared on exactly the same scale. $\mathrm F_{10.7cm}$ and the SN are both absolute indices. Indeed, they do not lean on external references, which anyway are absent over a large part of their temporal range (410 years for the SN, and 73 years for $\mathrm F_{10.7}$ ). Instead, the index values are only based on the knowledge of the different steps in their determination, from the raw measurements to the data processing method. For the SN, a full revision process was undertaken in 2011 and led to a first re-calibration of this multi-century series, with correction reaching up to 20\% \citep{{CletteEtal2014},{CletteLefevre2016}}. This re-calibration included a full re-construction of the SN from raw original data for the period 1981 to the present, while correction factors were applied to the original series built by the Z\"urich Observatory before 1981.

In this article, we will take a closer look at the $\mathrm F_{10.7cm}$ radio flux, which comes second in duration after the SN, among the global long-duration solar activity indices based on a single uninterrupted observing and processing technique, and on a single long-duration standard reference. Since the measurements of the $\mathrm F_{10.7cm}$ radio flux were started in Ottawa in 1947 \citep{{Covington1948}, {Covington1952}}, this new solar index proved to be highly correlated with the SN. This can be explained by the fact that the background flux, outside flaring events, is associated with the thermal free-free and gyroresonance emission of electrons trapped in closed loops anchored in the active regions, and thus primarily in their sunspots \citep{{Tapping1987}, {TappingDetracey1990}, {TappingZwaan2001}}. Those two indices ran in parallel over the last 73 years and they are produced by completely different and independent processes. Therefore, as they are supposed to retrace exactly the same evolution of the last 7 solar activity cycles, studying their mutual relation can give a prime diagnostic of their long-term stability. We will thus focus on the proxy relation between those two indices.

It turns out that, given the excellent long-term correlation between the SN and the $\mathrm F_{10.7cm}$ radio flux, various proxy relations were derived over past years by different authors, or they were established for operational purposes by solar data services like NOAA-SWPC (Space Weather Prediction Center) in the USA or the IPS (Ionospheric Prediction Service, part of the Bureau of Meteorology) in Australia. Those relations are motivated by two kinds of applications. One of them is the re-construction of the $F_{10.7}$ time series before the actual measurements started in 1947. Indeed, as the sunspot number extends back over four centuries, it allows to extrapolate this radio index over a much longer period \citep{Svalgaard2016}. Another application is producing mid-term predictions of the future evolution of the $\mathrm F_{10.7cm}$ flux. As such predictions often need to be calibrated and validated over many past solar cycles, they are typically based on the SN, and consequently, they produce their predictions in terms of this SN.  Moreover, the number of sunspots and sunspot groups give a direct measure of magnetic flux emergence at the solar surface, and is thus directly related to the dynamo mechanism at work inside the Sun \citep{{Charbonneau2010},{Hathaway2010},{Stenflo2012}}. 

On the other hand, $F_{10.7}$ is a chromospheric/coronal index that combines two kinds of emission: gyroresonance and free-free, the latter being associated with the magnetic decay of active regions under the action of the random convection, leading to the chromospheric plage component \citep{{Tapping1987}, {TappingZwaan2001}}. This is why, on timescales shorter than the average lifetime of individual active regions and their associated plages, daily values of $F_{10.7}$ are less correlated with the sunspot number, as first found by \cite{VitinskyPetrova1981}, \cite{Vitinsky1982}, \cite{Kopecky1982} and \cite{Kuklin1986}, and using more modern methods by \cite{Dudok2009} and \cite{DudokEtal2014}. On the other hand, the daily flux offers a better proxy for ultraviolet (UV) and X-ray fluxes produced in the chromosphere, the transition region and the solar corona. For this reason, $F_{10.7}$ is used by preference to the sunspot number for short-term forecasts of solar irradiance in the UV to X-ray domain and of its influence on the Earth environment (ionosphere, stratospheric temperatures, chemistry of the upper atmosphere), and for the resulting applications (radio propagation, atmospheric drag on low-Earth orbiting satellites). This close relation with the solar UV irradiance also allows to produce backward reconstructions of past UV fluxes for epochs well before the advent of direct space-based measurements of those fluxes \citep{Svalgaard2016}. 

However, beyond their strong Sun-related similarities on long timescales, the two indices differ by two base characteristics that play a role primarily at the lowest levels of activity. Firstly, $F_{10.7}$ does not fall to 0 when the Sun is spotless. A base background flux exists even when the Sun is fully quiet. This background emission, which corresponds to the spatially diffuse component of $F_{10.7}$, is probably associated with the small magnetic loops rooted in the quiet-Sun chromospheric network \citep{TappingZwaan2001}. This lower limit is still a matter of debate, but it is generally estimated in the range between 64 and 67 solar flux units (sfu)  \citep{{TappingDetracey1990},  {TappingCharrois1994}}.

Secondly, by its definition \citep{{Wolf1856},{CletteEtal2014}}, the SN is quantized at the lowest values, as each new group (with at least one single spot), adds 11 to the index. So, for the first spot, the SN jumps from 0 directly to 11. This effect quickly decreases for values larger than 22, as contributions from several groups with multiple sunspots are then combined in the total number. However, this low jump stretches the SN scale near 0. As this SN feature is absent in $F_{10.7}$, we can expect that it will break the proportionality between the two indices in the lowest range.

In this article, we first review all $F_{10.7}$\,--\,$S_{\mathrm{N}}$ proxy relations published in the literature or used by operational space weather services. Given the mismatches between those existing proxies, we build more carefully a new least-square polynomial regression, while exploring the effect of temporal averaging of the source data. We also investigate the issue of the quiet-Sun $F_{10.7}$ background level. We then check the temporal stability of the relation between $F_{10.7}$ and the SN over the entire duration of the series. We finally conclude on the new picture emerging from our analysis and on important aspects to be taken into account for future updates of this relation between those two most fundamental measures of the long-term solar activity.   

In this analysis, we use the sunspot number data provided by the World Data Center SILSO (Sunspot Index and Long-term Solar Observations) 
at \url{http://www.sidc.be/silso/datafiles}
and the $F_{10.7cm}$ data series from the Dominion Radio Astrophysical Observatory, available via the Space Weather Canada service at \url{https://www.spaceweather.gc.ca/solarflux/sx-5-en.php}, and also accessible through NOAA (\url{https://www.ngdc.noaa.gov/stp/space-weather/solar-data/solar-features/solar-radio/noontime-flux/penticton/}). We use the adjusted $F_{10.7cm}$ flux, which reduces the flux to a fixed distance of 1 Astronomical Unit (AU), and thus eliminates any annual modulation due to the orbital eccentricity of the Earth.

\section{Past $\mathrm F_{10.7cm}$\,--\,SN proxy relations}  \label{S:PastProxies}

\begin{figure}
	\centering
	\includegraphics[width=1.\columnwidth]{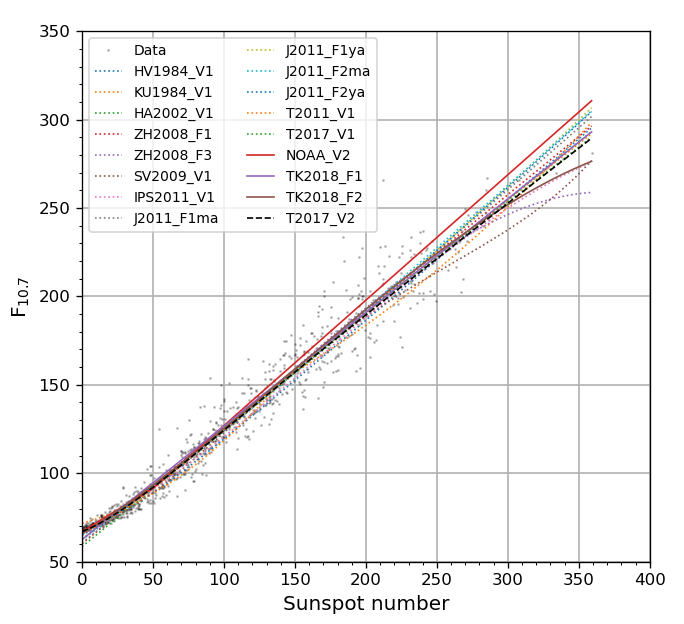}
		\caption{\small 
			Combined plot of past published proxy relations giving $F_{10.7}$ as a function of the sunspot number. The curves are labeled according to the identification in column 4 of Table \ref{T:ListProxies}. The curves are superimposed on the observed monthly mean values (gray dots).  \label{F:PastProx_full}} 
 	    
\end{figure}

Quite a number of proxy relations were proposed in the past. Here, we first compile all relations accessible in the literature or documented with associated data products at data centers 
\footnote{The Svalgaard (2009) proxy was found in the Web source: \url{https://wattsupwiththat.com/2009/05/18/why-the-swpc-10-7-radio-flux-graph-is-wrong/}. The NOAA-SWPC (2016) proxy is used for solar cycle predictions provided at	\url{https://www.swpc.noaa.gov/products/predicted-sunspot-number-and-radio-flux} and \url{https://www.swpc.noaa.gov/products/solar-cycle-progression}.	This proxy formula was formerly mentioned in an on-line document (\url{ftp://ftp.swpc.noaa.gov/pub/weekly/Predict.txt}), which is not accessible anymore. Some related information can presently be found in
\url{https://www.swpc.noaa.gov/sites/default/files/images/u2/Usr_guide.pdf}.}. 
They are listed in Table \ref{T:ListProxies}, and the corresponding formulae are given in Table \ref{T:ListProxFormulae}. In Figure \ref{F:PastProx_full}, we plot all those proxies together, superimposed on the monthly mean values of $F_{10.7}$ versus $S_{\mathrm{N}}$ version 2, the most recent re-calibration of this series \citep{CletteLefevre2016}. 

\begin{table}
		\caption{\small List of the proxies included in our comparison. For each proxy, we indicate the temporal granularity of the series used to fit the proxy (day, month, year; simple means or running means), and the SN version on which the proxy was based. The label identifies the curves in the associated figures, and the corresponding formulae are given in Table\,\ref{T:ListProxFormulae}. 
	    }
		\label{T:ListProxies}
	\centering
	\begin{tabular}{l l c l}
		\hline		
		Source & Temporal base & SN version & Plot label \\
		\hline
        \citet{Kuklin1984} & Unknown & 1 & KU1984\_V1 \\
		\citet{HollandVaughn1984} & 13-month smoothed & 1 & HV1984\_V1 \\
		\citet{XanthakisPoulakos1984} & 1 day & 1 & XP1985\_V1 \\
		\citet{HathawayEtal2002} & 24-month Gaussian smoothed & 1 & HA2002\_V1 \\
		\citet{ZhaoHan2008} formula 1  & 1-year mean & 1 & ZH2008\_F1 \\
		\citet{ZhaoHan2008} formula 3  & 1-year mean &
		1 & ZH2008\_F3 \\
		Svalgaard (2009)  & 1-month mean & 1 & SV2009\_V1 \\
		IPS Australia \citep{Thompson2010} & 1-month mean & 1 & IPS2011\_V1 \\
		\citet{TappingValdes2011} & 1-year mean & 1 & T2011\_V1 \\
		\citet{Johnson2011} formula 1 monthly & 1-month mean & 1 & J2011\_F1ma \\
		\citet{Johnson2011} formula 1 yearly & 1-year mean & 1 & J2011\_F1ya \\
		\citet{Johnson2011} formula 2 monthly & 1-month mean & 1 & J2011\_F2ma \\
		\citet{Johnson2011} formula 2 yearly & 1-year mean & 1 & J2011\_F2y \\
		NOAA-SWPC (2016)  & Unknown &	2 & NOAA\_V2 \\
		\citet{TappingMorgan2017} $S_{\mathrm{N}}$ version 1 & 10-month smoothed & 1 & T2017\_V1 \\
		\citet{TappingMorgan2017} $S_{\mathrm{N}}$ version 2 & 10-month smoothed & 2 & T2017\_V2 \\
		\citet{TiwariKumar2018} formula 1 & 1-month mean & 2 & TK2018\_F1 \\
		\citet{TiwariKumar2018} formula 2 & 1-month mean & 2 & TK2018\_F2 \\
		\hline
	\end{tabular}
	\newline
	\vspace{2mm}
	\newline
\end{table}

\begin{table}
		\caption{\small List of the proxies (labels from Table\,\ref{T:ListProxies}) and the corresponding formulae.}
		\label{T:ListProxFormulae}
	\centering
	\begin{tabular}{l l}
		\hline		
		Plot label & Formula \\
		\hline
		KU1984\_V1 & $71.74 + 0.2970 \,S_{\mathrm{N}} + 0.005146 \,S_{\mathrm{N}}^2 \, for \,S_{\mathrm{N}} < 100.5$, \\
		& $104.63 + 0.3037 \,S_{\mathrm{N}} + 0.001817 \,S_{\mathrm{N}}^2 \, for \,S_{\mathrm{N}} > 100.5$ \\
		HV1984\_V1 & $67.0 + 0.97 \, S_{\mathrm{N}} + 17.6 \, 
		(e^{-0.035 \, S_{\mathrm{N}}} - 1) $ \\
		XP1985\_V1 & $68.15 + 0.65\, S_{\mathrm{N}}$ \\
		HA2002\_V1 & $58.52 + 0.926\, S_{\mathrm{N}}$ \\
		ZH2008\_F1 & $60.1 + 0.932\, S_{\mathrm{N}}$ \\
		ZH2008\_F3 & $65.2 + 0.633\, S_{\mathrm{N}} + 3.76\, 10^{-3} S_{\mathrm{N}}^2 + 1.28\, 10^{-5} S_{\mathrm{N}}^3$ \\
		SV2009\_V1 & $67.29 + 0.316\, S_{\mathrm{N}} + 1.084\, 10^{-2} S_{\mathrm{N}}^2 + 6.813\, 10^{-5} S_{\mathrm{N}}^3  + 1.314\, 10^{-7} S_{\mathrm{N}}^4$ \\
		IPS2011\_V1 & $67.0 + 0.572\, S_{\mathrm{N}} + 3.31\, 10^{-3} S_{\mathrm{N}}^2 – 9.13\, 10^{-6} S_{\mathrm{N}}^3$ \\
		T2011\_V1 & $66 + 0.446\, S_{\mathrm{N}} (2. - e^{-0.027\,S_{\mathrm{N}}})$ \\
		J2011\_F1ma & $60.72 + 0.900\, S_{\mathrm{N}} + 0.0002\, S_{\mathrm{N}}^2$ \\
		J2011\_F1ya & $62.87 + 0.835\, S_{\mathrm{N}} + 0.0005\, S_{\mathrm{N}}^2$ \\
		J2011\_F2ma & $62.72 + (0.686\, S_{\mathrm{N}})^{1.0642}$ \\
		J2011\_F2y & $64.98 + (0.582\, S_{\mathrm{N}})^{1.0970}$ \\
		NOAA\_V2 & $67.00 + 0.4903\, S_{\mathrm{N}} \,for \,S_{\mathrm{N}} \le 50$; $56.06 + 0.7092\, S_{\mathrm{N}} \,for \,S_{\mathrm{N}} > 50$ \\
		T2017\_V1 & $67 + 0.44\, S_{\mathrm{N}}\,  [2. - e^{-0.031\, S_{\mathrm{N}}}]$ \\
		T2017\_V2 & $67 + 0.31\, S_{\mathrm{N}}\, [2. - e^{-0.019\, S_{\mathrm{N}}}]$ \\
		TK2018\_F1 & $62.51 + 0.6422\, S_{\mathrm{N}}$ \\
		TH2018\_F2 & $65.6605 + 0.500687\, S_{\mathrm{N}} + 1.21647\, 10^{-3} S_{\mathrm{N}}^2 + 2.71853\, 10^{-6}  S_{\mathrm{N}}^3$ \\
		\hline
	\end{tabular}

\end{table}
	

So far, only a few recent proxy relations were calculated for this new version of the SN series, $S_{\mathrm{Nv2}}$ released in July 2015. Therefore, for converting older proxies based on $S_{\mathrm{Nv1}}$ to the $S_{\mathrm{Nv2}}$ scale, we used the following re-scaling relation:
\begin{equation}
S_{\mathrm{Nv2}}= S_{\mathrm{Nv1}} / 0.6 / 1.177
\label{E:Snv1SnV2}
\end{equation}

\begin{figure}
\centering
\includegraphics[width=1.0\columnwidth]{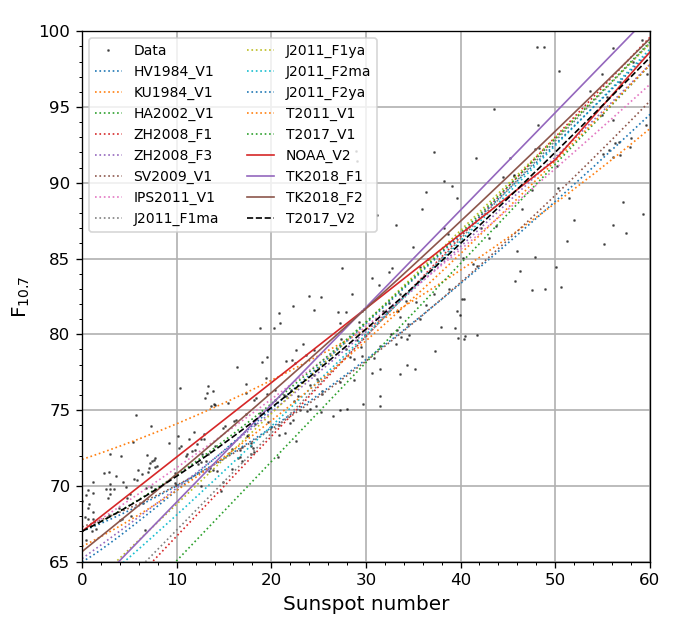}
	\caption{\small 
	Combined plot of past published proxy relations giving $F_{10.7}$ as a function of the sunspot number: close-up view of the low-activity range of Figure \ref{F:PastProx_full}. The curves are labeled according to the identification in column 4 of Table \ref{T:ListProxies}. The curves are superimposed on the observed monthly mean values (gray dots). \label{F:PastProx_zoom}} 
\end{figure}

The 0.6 factor is associated with a change of reference observer, making the former 0.6 conventional Z{\"u}rich factor obsolete. The second factor corresponds to a correction for an artificial inflation of the original SN values due to the use of a weighting according to the sunspot size artificially introduced in Z{\"u}rich \citep{CletteEtal2016}. This was affecting all the $S_{\mathrm{Nv1}}$ values after 1946. As the $F_{10.7}$ series starts in 1947, this relation is thus valid for the entire time interval considered here. The 1.177 factor is in fact an asymptotic values reached for medium to high levels of solar activity. It is thus variable at low solar activity, dropping to almost 1 near cycle minima. Therefore, Equation \ref{E:Snv1SnV2} is not fully accurate for low-activity values. Still, in our analysis presented below, we did not find any significant deviation between version 1 and version 2 proxies due to this effect, at the level of precision associated with the data themselves. This can be seen in the closeup view (Fig. \ref{F:PastProx_zoom}).

Overall, we observe that some proxies are very crude. They are simple linear fits, ignoring the visible deviation from linearity at the lower end of the range. Strong deviations also appear at the high values, in particular for non-linear fits (polynomial or exponential models). This can be explained by the limited number of such high values in the past solar activity record, which thus leads to large uncertainties. We also note that in the low range below $S_{\mathrm{N}}=20$, virtually all proxies fall below the observed values, and thus lead to systematic underestimates of the average $F_{10.7}$ flux at low activity. 

\begin{figure}
	\centering
	\subfigure{\includegraphics[width=0.7\columnwidth,trim=0cm 0.2cm 0cm 0.5cm,clip]{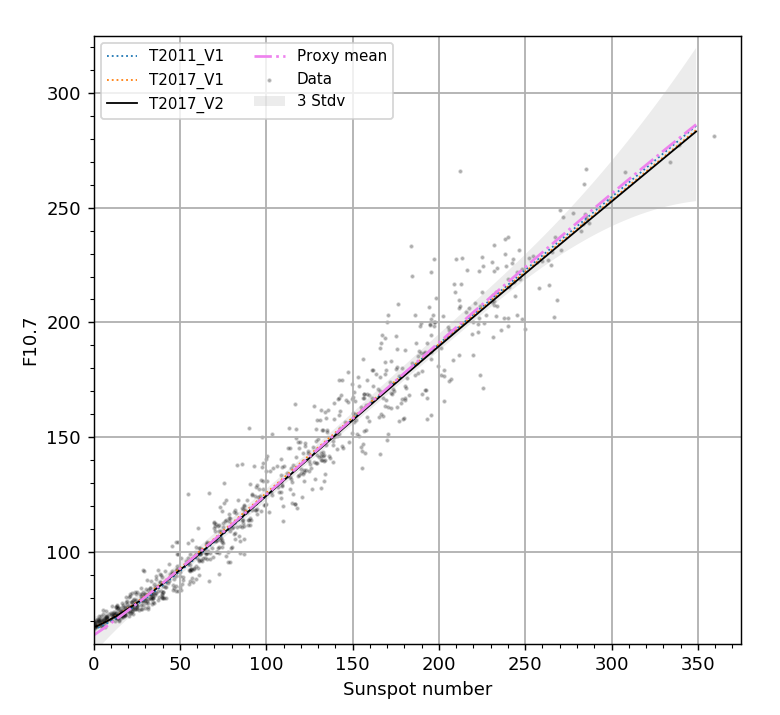}}
	\subfigure{\includegraphics[width=0.7\columnwidth,trim=0cm 0.2cm 0cm 0.5cm,clip]{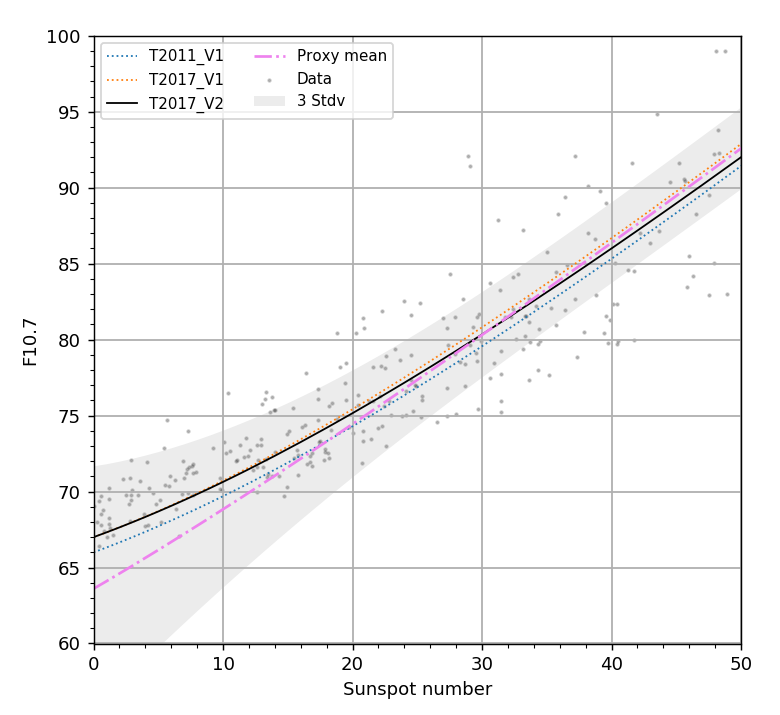}}
		\caption{\small 
			Mean (purple dash-dotted curve) and standard deviation of all proxies in Figure \ref{F:PastProx_full} (gray shading). Three recent proxies by Tapping et al. (cf. Table \ref{T:ListProxies}) are also included, as well as the observed monthly mean values (gray dots). The lower plot is a zoomed-in view of the upper plot for low activity levels. \label{F:PastProx_mean}} 
\end{figure}

Moreover, some proxies were derived using monthly means or yearly means. In that case, the upper range of values is more limited, and the fits should not be trusted beyond their calibration range. Unfortunately, while a few estimates of the error of individual daily $F_{10.7}$ flux values were published \citep{NicoletBossy1985,TappingCharrois1994}, we must note that most of the available proxy relations are given without any estimate of their uncertainties, and often without clear indication of the calibration range. Here, we conservatively derived the mean and standard deviation of all proxy models shown in the plot (black line and shaded band in Fig. \ref{F:PastProx_mean}), to get a rough first idea of their actual uncertainty.

The proxies giving the best fit to the non-linear section at low SN are those published by \citet{TappingValdes2011} (based on $S_{\mathrm{Nv1}}$) and \citet{TappingMorgan2017} (based on $S_{\mathrm{Nv1}}$ and $S_{\mathrm{Nv2}}$). The authors mention that those proxies were defined purely empirically, and they do not explain how they were adjusted on the data. Both proxies are shown in Figure \ref{F:PastProx_mean}. One can see that the $S_{\mathrm{Nv1}}$ and $S_{\mathrm{Nv2}}$ proxies are almost identical, indicating that the conversion in Equation[1] is accurate. However, we note that all those proxies reach a value of 67 sfu for $S_{\mathrm{N}}=0$, while almost all observed $F_{10.7}$ values are above this lower limit. In fact, below in Section \ref{S:BackgndFlux}, we find that the most probable $F_{10.7}$ value for a spotless Sun is 70.5 sfu. This mismatch indicates that this part of the curve was not derived by least-squares but was adjusted empirically to reach exactly a tie-point at 67 sfu, chosen as base quiet-Sun background when $S_{\mathrm{N} = 0}$. Given the mismatch with the actual data, this choice seems questionable. Indeed, for real applications, users need the most probable $F_{10.7}$ flux, and not the lowest possible value, which is rarely reached. In Section \ref{S:BackgndFlux}, we will consider more closely the properties of this quiet-Sun $F_{10.7}$ background. 

Still, the other published proxies are underestimating even more the $F_{10.7}$ flux at low activity. \emph{Therefore, overall, none of the proxies proposed so far are providing a satisfactory representation of the relation at low solar activity.} Moreover, we note that all past proxies used classical least-square fits, which assume that errors are present only in the fitted measurement (here $F_{10.7}$), while the other quantity ($S_{\mathrm{N}}$) is considered as a parameter (without error). As $S_{\mathrm{N}}$  is also affected by errors, this fitting model may thus lead to systematic biases. We also checked this aspect as explained in sub-Section \ref{SS:PolynODR} below.

\section{Mean profiles} \label{S:MeanProf}
In order to extract the $F_{10.7}/S_{\mathrm{N}}$ relation without any parametric model, we first derive the mean of $F_{10.7}$ values and $\sigma_m$, the standard error of the mean (SEM), for a given value of $S_{\mathrm{N}}$. As the temporal averaging of raw daily values will influence the relation between the two quantities, we repeated this calculation for raw daily value pairs, for monthly means, for 13-month smoothed monthly values and for yearly means. In order to include a sufficiently large sample of values and to reduce random noise effects, we derived the statistics over a limited $S_{\mathrm{N}}$ range centered on each given $S_{\mathrm{N}}$ value. The bin width was 3, 20, 20, 60 respectively for daily, monthly, smoothed and yearly values. As our analysis immediately showed that results for yearly means and 13-month smoothed data are almost identical, we will not further discuss the 13-month smoothed results here. 

\begin{figure}
	\centering
	\includegraphics[width=1.\columnwidth]{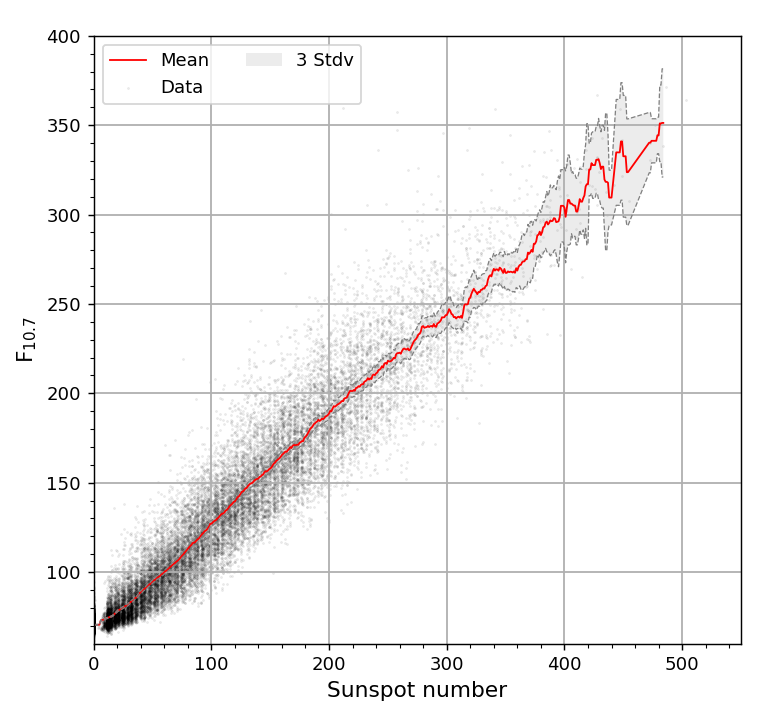}
		\caption{\small 
			Mean non-parametric profile (red line) with $3\,\sigma_m$  range (gray shading), obtained by averaging all daily $F_{10.7}$ values for $S_{\mathrm{N}}$ within a narrow band centered on the $S_{\mathrm{N}}$ in the bottom axis (See details in the main text). Gray dots are the daily observed values. The number of data points strongly decreases above $S_{\mathrm{N}}=300$ and $F_{10.7}=250$ sfu, leading to a much larger SEM $\sigma_m$. 
		\label{F:MeanCurve_d}} 

\end{figure}

\begin{figure}
	\centering
	\includegraphics[width=1.\columnwidth]{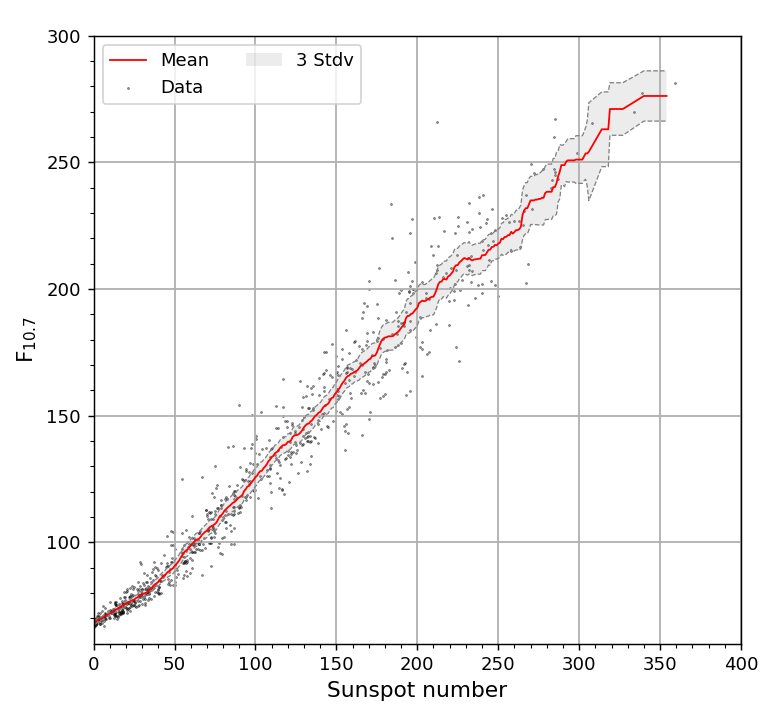}
		\caption{\small 
			Mean non-parametric profile (red line) with $3\,\sigma_m$ range (gray shading), like in Figure \ref{F:MeanCurve_d} but obtained by averaging monthly mean $F_{10.7}$ values over narrow $S_{\mathrm{N}}$ bands. Gray dots are the monthly mean observed values. \label{F:MeanCurve_m}} 
\end{figure}

\begin{figure}
	\centering
	\includegraphics[width=1.\columnwidth]{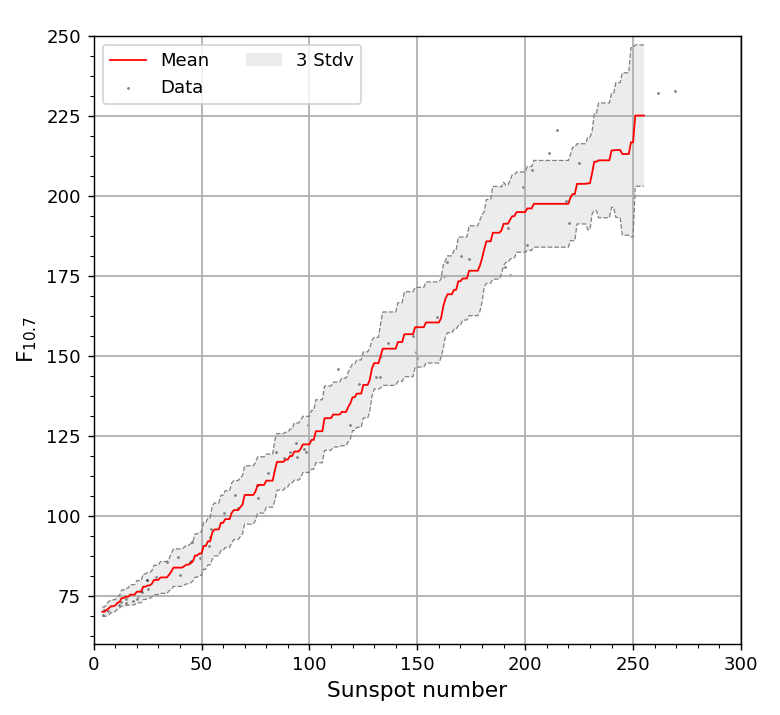}
		\caption{\small 
			Mean non-parametric profile (red line) with $3\,\sigma_m$  range (grey shading), like in Figure \ref{F:MeanCurve_d} but obtained by averaging yearly mean $F_{10.7}$ values over narrow $S_{\mathrm{N}}$ bands. Gray dots are the yearly mean observed values.	\label{F:MeanCurve_y}} 
\end{figure}

Figures \ref{F:MeanCurve_d}, \ref{F:MeanCurve_m} and \ref{F:MeanCurve_y} show the resulting mean curves and 3-$\sigma_m$ band for the daily, monthly, and yearly calculations. While the standard deviation of the base daily data is quite large, in particular for raw daily values (16.7\,sfu overall), the SEM value is rather small in the low and medium range ($< 1$\,sfu for daily values), thanks to the fact that each mean value is based on a large number of data points within each $S_{\mathrm{N}}$ slice (about 400 points). It strongly increases in the upper range, above $S_{\mathrm{N}}=200$, as data points become sparse, indicating that the fits will become much less precise in that upper range. 

\begin{figure}
	\centering
 	\includegraphics[width=1.\columnwidth]{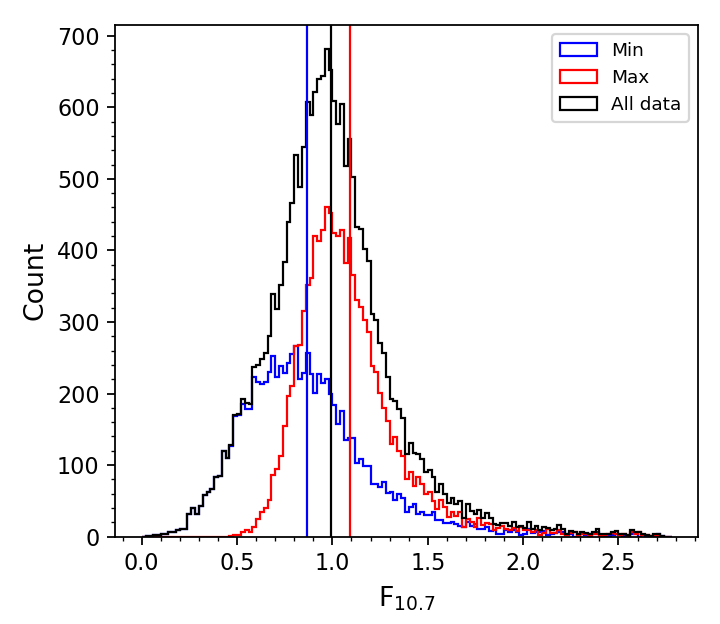}
		\caption{\small 
			Histogram of the ratio between daily $F_{10.7}$ data and the mean $F_{10.7}$ value (after subtraction of a 67 sfu base quiet-Sun flux), for all data with $S_{\mathrm{N}} > 15$. The black curve is for the entire data series, while the blue and red histograms are respectively for the maxima of the solar cycles (time of maximum -2 to +3 years) and the minima (the rest of the data). The distributions are slightly asymmetrical with a longer upper wing. The means of the distributions are indicated by thick vertical lines, with the matching colors. The standard deviation of the distributions equals 35\%, while the SEM equals 0.5\% (based on more than 10000 daily ratios in each distribution). The distributions for cycle maxima and minima are significantly shifted by 12\% above and below the global mean. 	\label{F:Hist_d}} 
\end{figure}

Now considering the raw daily values, we observe that the distribution of individual $F_{10.7}$ values around the mean is asymmetrical with a more extended wing towards high values (Figure \ref{F:Hist_d}).This upper wing may result from different contributions. This excess flux may come from the incomplete elimination of eruptive events and from the temporal under-sampling of the 20h00UT ''spot'' measurements, as indicated by \citet{TappingCharrois1994}. If any flaring emission was present over time intervals when the S-component background emission was extracted, it inevitably contributed to an overestimate of the background flux and thus to a net positive excess, contrary to simple random measurement noise. This thus leads to an upwards asymmetry of the random deviations, like we find here. 

Another consequence of this deviation from the assumed symmetrical Gaussian noise distribution should be a small upward bias in the estimate of the mean. Moreover, when considering temporal variations (see Section \ref{S:TempVar}), we find that the distribution is also slightly higher around the maxima of solar cycles (time of maximum -2 years to +3 years; blue curve in Figure \ref{F:Hist_d}) than around solar minima (red curve), by about 12\%. This thus suggests a significant change in the $F_{10.7}$ -- $S_{\mathrm{N}}$ relation over each solar cycle. We will further examine this interesting property in Section \ref{S:TempVar}.

However, this min -- max shift, as well as the asymmetry of the distribution strongly decreases when considering longer time scales, i.e. for monthly and yearly means. The distribution becomes Gaussian and constant over time over those longer timescales. We can observe this by comparing daily data with the plots for the monthly and yearly means (Figures \ref{F:MeanCurve_m} and \ref{F:MeanCurve_y}). We find that all temporally averaged values lead to very consistent means, and that those means are all slightly lower than the mean for daily values (though still marginally consistent within the SEM $\sigma_m$). This indicates that the upward bias affecting raw daily values largely vanishes for longer time scales. Moreover, $\sigma_m$ values are lower, and similar for the 13-month smoothed and yearly means, as expected: 10.0 sfu (monthly), 6.2 sfu (smoothed), 6.3 sfu (yearly). This lower dispersion and higher correlation between $F_{10.7}$ and the SN indicates that for long duration, beyond a single solar rotation, both indices record the level of solar activity (total magnetic flux emergence) essentially in the same way. The global emergence rate dominates the statistics, and the skewed randomness of the $F_{10.7}$ flux on short times scales only plays a minor role. We will deepen this interpretation in Section \ref{S:TempVar}.

In all cases, we can see that over most of the observed range, the means trace a largely linear proportionality between the two indices. Only above $S_{\mathrm{N}}=250$, $F_{10.7}$ tends to fall slightly below the linear relation. This slight curvature would suggest that for short periods of extreme activity during cycle maxima, $F_{10.7}$ does not grow as fast relative to the sunspot number. However, taking into account the large uncertainty, this is only marginally significant, and a fully linear relation remains valid up to the highest observed $F_{10.7}$ fluxes. 

\begin{figure}
	\centering
	\includegraphics[width=1.\columnwidth]{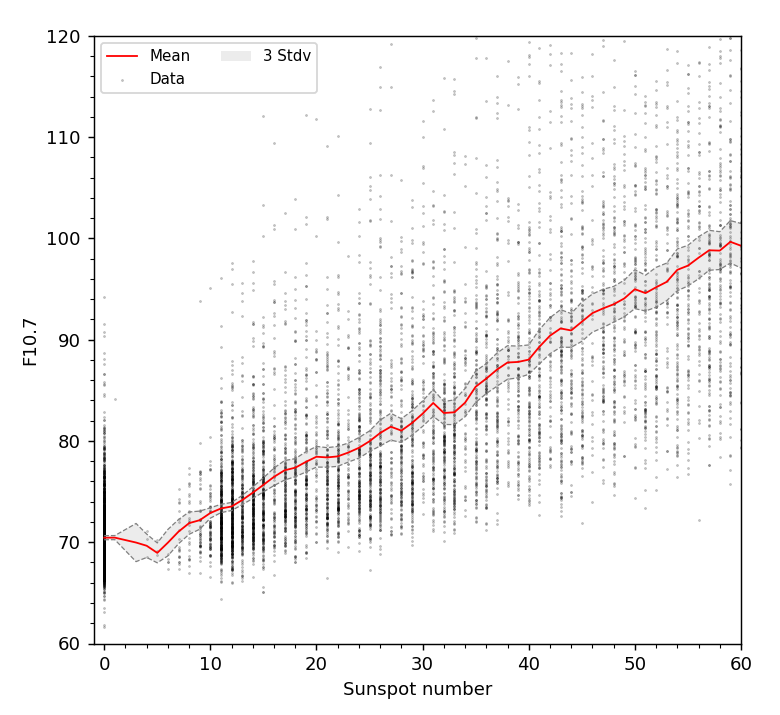}
		\caption{\small 
			Mean non-parametric profile (red line) with $3\,\sigma_m$ range (grey shading) for daily values: enlarged view of the low activity part of Figure \ref{F:MeanCurve_d}.	\label{F:MeanCurve_dz}} 
\end{figure}

\begin{figure}
	\centering
	\includegraphics[width=1.\columnwidth]{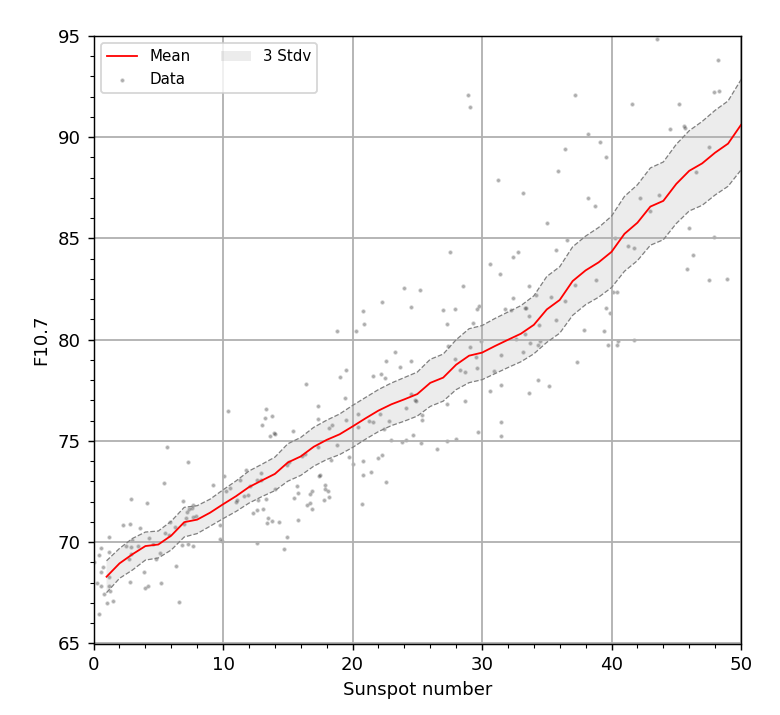}
		\caption{\small 
			Mean non-parametric profile (red line) with $3\,\sigma_m$ range (grey shading) for monthly means: enlarged view of the low activity part of Figure \ref{F:MeanCurve_m}.	\label{F:MeanCurve_mz}} 
\end{figure}

\begin{figure}
	\centering
	\includegraphics[width=1.\columnwidth]{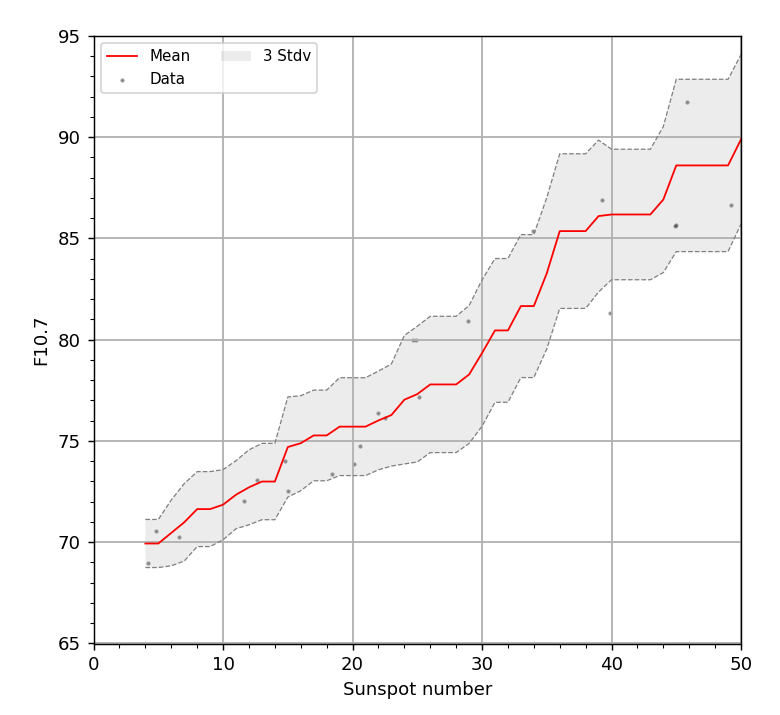}
		\caption{\small 
			Mean non-parametric profile (red line) with $3\,\sigma_m$ range (grey shading) for yearly means: enlarged view of the low activity part of Figure \ref{F:MeanCurve_y}.	\label{F:MeanCurve_yz}} 
\end{figure}

When zooming in on the low values (Figures\,\ref{F:MeanCurve_dz}, \ref{F:MeanCurve_mz}, \ref{F:MeanCurve_yz}), we find that the means for monthly, smoothed and yearly data start to deviate from the main linear part for $S_{\mathrm{N}}$ values below an inflection point at about $S_{\mathrm{N}}=35$ (monthly values) or 50 (smoothed and yearly values). As there are almost no monthly or yearly periods with a 0 mean $S_{\mathrm{N}}$,  the ordinate at $S_{\mathrm{N}}=0$ can only be extrapolated, and cannot be trusted. Only for monthly means, we find that the means tend towards 68 sfu for SN below 5.

On the other hand, for raw daily values, we find that the means continue to follow linear proportionality down to very low values, around $S_{\mathrm{N}}=5$.  There are only a very few points between 11 and 0, but at $S_{\mathrm{N}}=0$ , the mean value is well defined at 70.5 sfu. We find that the means reach this values within the uncertainties for $S_{\mathrm{N}}=8$, which is very low.  

We can thus draw three important conclusions:
\begin{itemize}
	\item \emph{$F_{10.7}$ and the SN are fully proportional over the full range of observed values, and this proportionality continues down to an almost spotless Sun.} This suggests that the variability of $F_{10.7}$ is entirely determined by the level of magnetic activity also controlling the number of sunspots, without any other contribution to the radio flux. The excess associated with the presence of spots becomes negligible relative to the $F_{10.7}$ flux distribution for a fully spotless Sun only below $S_{\mathrm{N}} = 8$. In other words, the distribution of the $F_{10.7}$ is largely the same for a fully spotless Sun and when a single isolated and short-lived spot is present. On the other hand, once the number of spots grows beyond one (one group with a single spot), $F_{10.7}$ increases fully proportionally with the sunspot number. So, in that sense, the quantization of the SN at low activity (the 0 to 11 jump) does not lead to a significant non-linearity between the two quantities, except for the values at $S_{\mathrm{N}}=0$. 
	\item \emph {As in temporal means, solar activity varies during the chosen time interval, the linear relation will be changed near the origin, essentially because of this single deviating point at $S_{\mathrm{N}}=0$.} As the latter is above the overall linear trend, the means will be pushed upwards, and this effect will increase as the mean SN decreases towards 0. Indeed, the time interval used for each mean will contain a growing proportion of spotless days, dominated by the $F_{10.7}$ = 70.5 sfu background. This is exactly what we find in monthly, smoothed and yearly means, with the non-linearity extending progressively to higher minimum SN as the duration of the temporal averaging increases. In Section \ref{S:DataModel}, we will build a simulation to validate this interpretation. 
	\item \emph{The proxy relation is consistent with Gaussian statistics and is stable only for timescales equal to or longer than one month.} Daily data and short timescales include an excess of high fluxes, which varies with the solar cycle. Those data are thus inappropriate for building a reliable proxy relation.
\end{itemize}

\section{New high-degree polynomial fits} \label{S:NewPolyn}

\subsection{Ordinary least-square polynomial fit: monthly means} \label{SS:PolynMonth}

In order to obtain a better fit to the data than the earlier, sometimes very crude, fits shown in Section \ref{S:PastProxies}, we fitted polynomials with degrees up to 4 by least-square regression of $F_{10.7}$ versus $S_{\mathrm{N}}$. Indeed, the non-parametric mean curves shown in the previous section indicate that the actual relation is largely linear over a wide range, with a rather sharp bifurcation towards a constant background in the low range. The polynomials are of the form (here for a $4^{th}$ degree polynomial):

\begin{equation}
F_{10.7}= C_0 + C_1\, S_{\mathrm{N}} + C_2\, S_{\mathrm{N}}^2 + C_3\,
 S_{\mathrm{N}}^3 + C_4\, S_{\mathrm{N}}^4
\label{E:Polyn}
\end{equation}

\begin{figure}
	\centering
	\subfigure{\includegraphics[width=0.7\columnwidth,trim=0cm 0.2cm 0cm 0.5cm,clip]{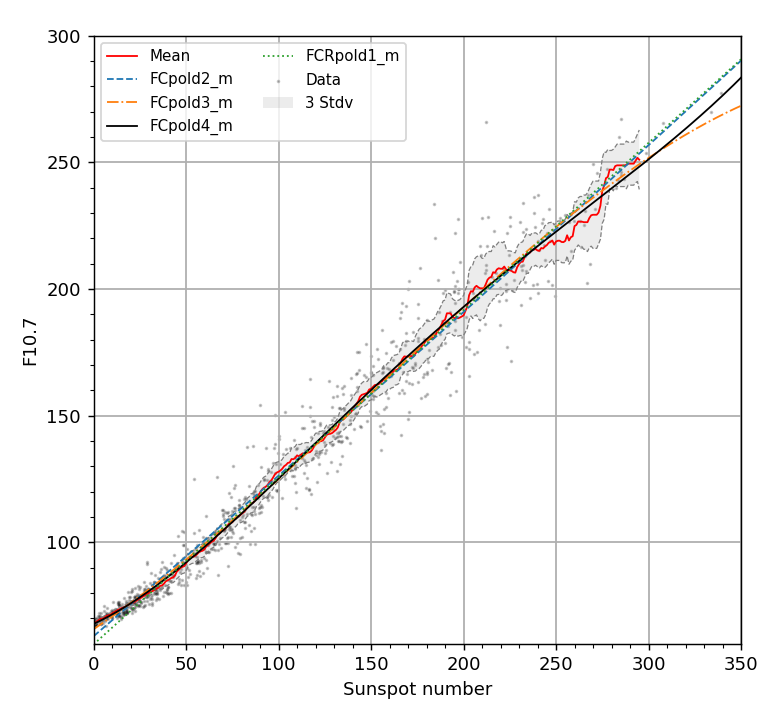}}
	\subfigure{\includegraphics[width=0.7\columnwidth,trim=0cm 0.2cm 0cm 0.5cm,clip]{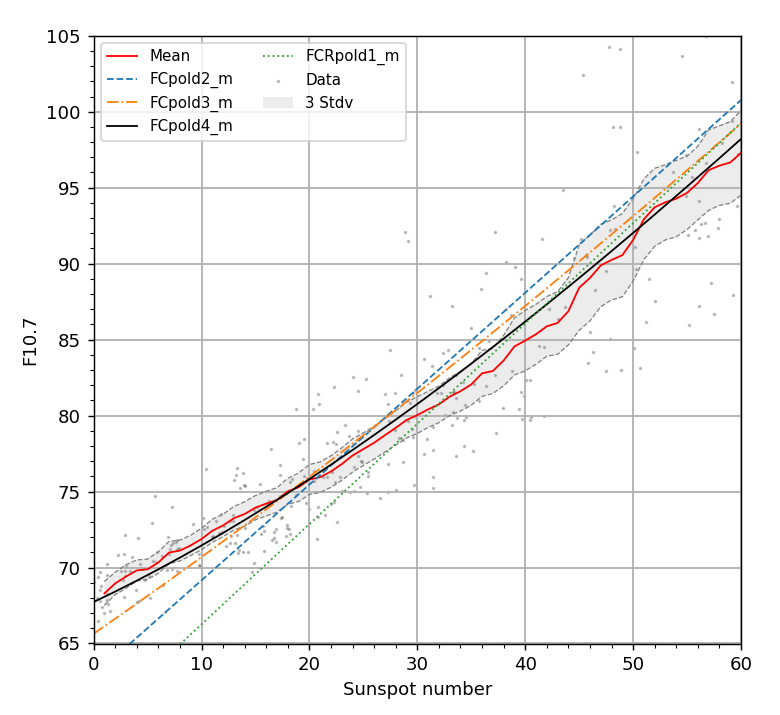}}
		\caption{\small 
		Polynomial of order 1 (linear fit) to 4 fitted to the monthly mean data by ordinary least-square regression (OLS). The curves are superimposed on the corresponding non-parametric mean (cf. Fig. \ref{F:MeanCurve_m}) to show the agreement within $3\,\sigma_m$, and on the base data (gray dots). The lower plot is a close-up view of the low range of the upper plot.	\label{F:Polyn1-4_m}} 
\end{figure}

In Figure \ref{F:Polyn1-4_m}, we show the fits to the monthly mean values for degrees 1 (linear) to 4. As we know that the relation becomes strongly non-linear below $S_{\mathrm{N}}=25$ , the linear fit (degree 1) was applied to a restricted range without the interval $S_{\mathrm{N}}=0$ to 25. So, this fit gives a good  model for the main linear section.

From $S_{\mathrm{N}}= 30$ to 250, all fits are almost identical and remain within the uncertainty range of the mean values (gray shaded band). Only above 250, there is a slight deviation, with the higher degrees falling below the linear fit. But this is hardly significant, given the low number of data points in this upper range. This is confirmed by the fact that coefficients of the high-degree terms are only marginally significant. In particular for the degree-2 polynomial, only the linear term (degree 1) is significant. This curve indeed gives the worst fit to the data.

This can be seen in the close-up of the low part (second plot), which shows that the fitted curve match progressively better the curved lowest part of as the polynomial degree increases. Only the 4th-degree polynomial closely reproduces the low part and remains within the uncertainty of the mean values. The coefficients for fitted polynomials up to degree 4 are listed in Table \ref{T:PolyOLSm}.

\begin{table}
  \caption{\small Coefficients of polynomials of order 1 to 4 fitted by the ordinary least-square regression on the monthly mean values. The coefficients $C_n$ correspond to Equation \ref{E:Polyn}, with their standard error $\sigma_n$. }
  \label{T:PolyOLSm}
  \centering
  \begin{tabular}{lllll}
	\hline	
Coefficients &  Order 1    &  Order 2   &  Order 3    &  Order 4    \\
             & (FCpol1\_m) &(FCpol2\_m) & (FCpol3\_m) & (FCpol4\_m) \\
	\hline	
$C_0$      & $62.31$  & $62.87$  &  $65.64$  &  $67.73$  \\
$\sigma_0$ & $0.5743$ & $0.7692$ &  $0.9457$ &  $1.134$  \\
$C_1$      & $0.6432$ & $0.6279$ &  $0.4918$ &  $0.3368$ \\
$\sigma_1$ & $4.528\,10^{-3}$ & $1.478\,10^{-2}$ &  $3.132\,10^{-2}$ &
  $5.649\,10^{-2}$ \\
$C_2$      &                  & $6.141\,10^{-5}$ &  $1.304\,10^{-3}$ &
  $3.690\,10^{-3}$ \\
$\sigma_2$ &                  & $5.637\,10^{-5}$ &  $2.592\,10^{-4}$ & 
 $7.699\,10^{-4}$ \\
$C_3$      &                  &                  & $-2.919\,10^{-6}$ &  
 $-1.517\,10^{-5}$ \\ 
$\sigma_3$ &                  &                  &  $5.946\,10^{-7}$ & 
 $3.773\,10^{-6}$ \\
$C_4$      &                  &                  &                   & 
 $1.974\,10^{-8}$ \\
$\sigma_4$ &                  &                  &                   &
 $6.003\,10^{-9}$ \\
    \hline	
    \end{tabular}
\end{table}

As the relation between the two indices is fully linear over a wide range, in this case above $S_{\mathrm{N}}=25$, we also derived the linear fits to this linear section. The coefficients are given in Table \ref{T:LinRegm} (also for the orthogonal regression method described below in sub-Section \ref{SS:PolynODR}). The linearity is confirmed by the fact that polynomials fits above degree 1 do not give stable solutions, once the lowest range is excluded (terms of degrees above 1 are not significant).

\begin{table}
	\caption{\small Coefficients of the linear fits to the monthly mean data in the restricted linear range $S_{\mathrm{N}}=25-290$ by ordinary least-squares and by orthogonal distance regression. The two fits match closely.}
	\label{T:LinRegm}
	\centering
	\begin{tabular}{lll}
		\hline	
Coefficients & Order 1          & Order 1 (ODR)    \\ 
             & (FCRpol1\_m)     &                  \\ 
$C_0$        & $59.66$          & $58.21$          \\
$\sigma_0$   & $0.8801$         & $0.8831$         \\
$C_1$        & $0.6601$         & $0.6720$         \\
$\sigma_1$   & $6.313\,10^{-3}$ & $6.338\,10^{-3}$ \\
		\hline	
	\end{tabular}
\end{table}

\subsection{Polynomial fits to yearly means} \label{SS:PolynYear}

\begin{figure}
	\centering
	\subfigure{\includegraphics[width=0.7\columnwidth,trim=0cm 0.2cm 0cm 0.5cm,clip]{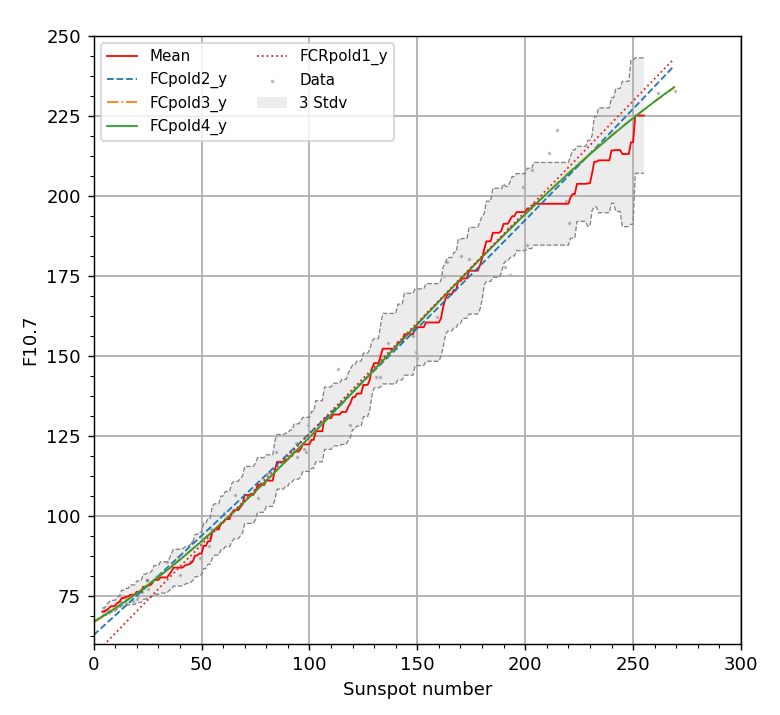}}
	\subfigure{\includegraphics[width=0.7\columnwidth,trim=0cm 0.2cm 0cm 0.5cm,clip]{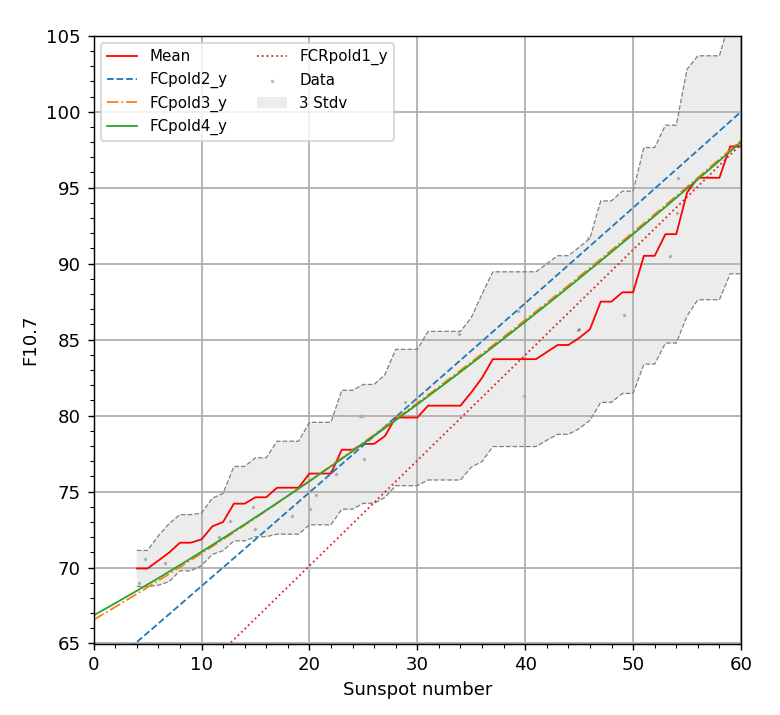}}
		\caption{\small 
			Polynomial of order 1 (linear fit) to 4 fitted to the yearly mean data by ordinary least-square regression (OLS). The curves are superimposed on the corresponding non-parametric mean (cf. Fig. \ref{F:MeanCurve_y}) to show the agreement within $3\,\sigma_m$, and on the base data (gray dots). The lower plot is a close-up view of the low range of the upper plot. \label{F:Polyn1-4_y}} 
\end{figure}

We repeated the analysis on yearly values and found largely the same conclusions. The curves are shown  in Figure \ref{F:Polyn1-4_y}, and the polynomial coefficients for the fits to yearly means are given in Tables \ref{T:PolyOLSy} and \ref{T:LinRegy}.

In this case, the fit is also not significant at polynomial order 2, and the order-4 polynomial gives roughly the same quality of fit as order 3. Although the fits on yearly means are slightly different from the fits derived from monthly mean values, both are compatible within the uncertainties in yearly means. This difference is due to a lower non-linearity and the slightly wider range over which the relation is non-linear for yearly means, but is hardly significant.

\begin{table}
	\caption{\small Coefficients of polynomials of order 1 to 4 fitted by the ordinary least-square regression on the yearly mean values. For the order 2 and 4 polynomials, coefficients for degree 2 and above are not significant (marked in {\it italics}), indicating that the proxy relation is essentially linear.}
	\label{T:PolyOLSy}
	\centering
	\begin{tabular}{lllll}
		\hline	
		Coefficients &  Order 1    &  Order 2   &  Order 3    &  Order 4    \\
		& (FCpol1\_y) &(FCpol2\_y) & (FCpol3\_y) & (FCpol4\_y) \\
		\hline	
		$C_0$      & $61.07$  & $62.64$  &  $66.56$  &  $66.85$  \\
		$\sigma_0$ & $ 1.323$ & $1.873$  &  $2.404$  &  $3.220$  \\
		$C_1$      & $0.6555$ & $0.6114$ &  $0.4163$ &  $0.3942$ \\
		$\sigma_1$ & $1.062\,10^{-2}$ & $3.893\,10^{-2}$ &  $8.761\,10^{-2}$ &
		$1.816\,10^{-1}$ \\
		$C_2$      &                  & $\it 1.862\,10^{-4}$ &  $2.129\,10^{-3}$ &
		$\it 2.510\,10^{-3}$ \\
		$\sigma_2$ &                  & $\it 1.583\,10^{-4}$ &  $8.031\,10^{-4}$ & 
		$\it 2.849\,10^{-3}$ \\
		$C_3$      &                  &                  & $-5.081\,10^{-6}$ &  
		$\it -7.327\,10^{-6}$ \\ 
		$\sigma_3$ &                  &                  &  $2.062\,10^{-6}$ & 
		$\it1.623\,10^{-5}$ \\
		$C_4$      &                  &                  &                   & 
		$\it 4.216\,10^{-9}$ \\
		$\sigma_4$ &                  &                  &                   &
		$\it 3.021\,10^{-8}$ \\
		\hline	
	\end{tabular}
\end{table}

\begin{table}
	\caption{\small Coefficients of the linear fits to the yearly mean data in the restricted linear range $S_{\mathrm{N}}= 30$ - 220 by ordinary least-squares.}
	\label{T:LinRegy}
	\centering
	\begin{tabular}{lll}
		\hline	
		Coefficients   & Order 1           \\ 
		               & (FCRpol1\_y)      \\ 
		$C_0$          & $56.24$           \\
		$\sigma_0$     & $2.259$           \\
		$C_1$          & $0.6936$          \\
		$\sigma_1$     & $1.739\,10^{-2}$  \\
		\hline	
	\end{tabular}
\end{table}

\subsection{Polynomial fits to daily values}

\begin{figure}
	\centering
	\subfigure{\includegraphics[width=0.7\columnwidth,trim=0cm 0.2cm 0cm 0.5cm,clip]{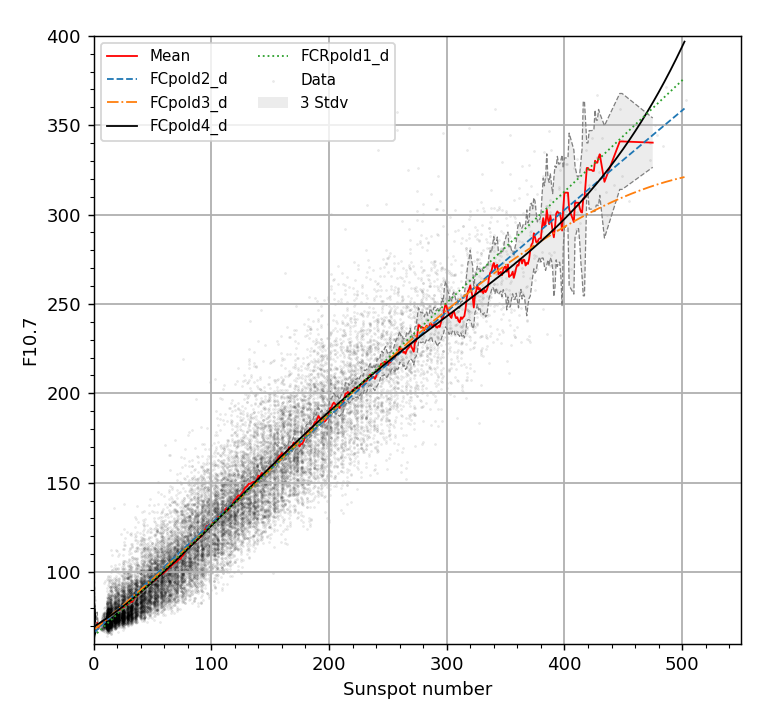}}
	\subfigure{\includegraphics[width=0.7\columnwidth,trim=0cm 0.2cm 0cm 0.5cm,clip]{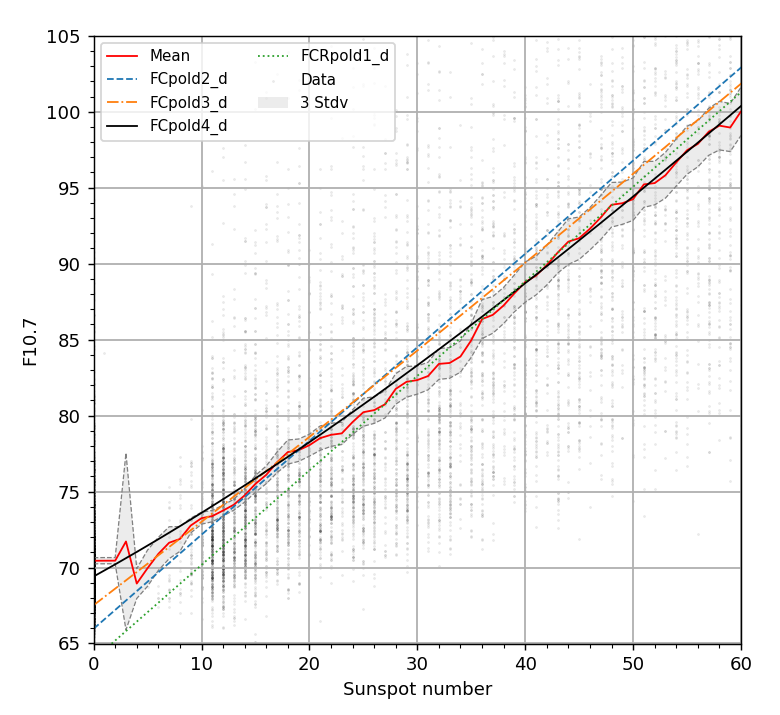}}
		\caption{\small 
			Polynomial of order 1 (linear fit) to 4 fitted to the daily data by ordinary least-square regression (OLS). The curves are superimposed on the corresponding non-parametric mean (cf. Fig. \ref{F:MeanCurve_d}) to show the agreement within $3\,\sigma_m$, and on the base data (gray dots). The lower plot is a close-up view of the low range of the upper plot. \label{F:Polyn1-4_d}} 
\end{figure}

Based on Section \ref{S:MeanProf}, we may expect slightly different results for daily values. The polynomial and linear regressions are shown in Figure \ref{F:Polyn1-4_d} and the coefficients are listed in Tables \ref{T:PolyOLSd} and \ref{T:LinRegd}. We find that for the main linear range up to $S_{\mathrm{N}}=250$, the different fits match closely, like for yearly and monthly means. They then diverge from each other at higher values, which is again due to the steeply decreasing number of data points in this upper range. The non-linear fits at order 2 to 4 tend to fall below the linear fit, aligning better with the mean values. However, the linear fit still falls within the uncertainty range. So, the differences brought by higher degrees are not fully significant. This is again confirmed by the low level of significance of polynomial coefficients with degrees higher than 1.

The low range is where the situation differs markedly from the monthly and yearly mean analyses. Here (Fig. \ref{F:Polyn1-4_d}), the linear fit (over the range above $S_{\mathrm{N}}=25$) remains close to the mean values down to the lowest values. In the low range, the order-4 curve again gives the best fit, although it fails for $S_{\mathrm{N}}$ values below 6, as the mean then deviates abruptly over a very small range. Ignoring this section, the order-3 polynomial gives the best fit overall. It reaches the mean background value for $S_{\mathrm{N}}=0$ (70.5 sfu) at $S_{\mathrm{N}}=5$.  This suggests that for $S_{\mathrm{N}}$ below 6, this background value can be used instead of the polynomial fit.

Therefore, the fit on daily values helps to confirm the higher linearity of the $F_{10.7}$ -- $S_{\mathrm{N}}$ relation down to very low levels. However, as expected, the fitted curves are significantly higher than the fits on monthly and yearly means, due to the upward bias characterizing raw daily values (see Section \ref{S:MeanProf}). Beyond the linearity check, they should thus not be considered for a proxy relation.  

\begin{table}
	\caption{\small Coefficients of polynomials of order 1 to 4 fitted by the ordinary least-square regression on the daily values. The coefficients $C_n$ correspond to Equation \ref{E:Polyn}, with their standard error $\sigma_n$. }
	\label{T:PolyOLSd}
	\centering
	\begin{tabular}{lllll}
		\hline	
		Coefficients &  Order 1    &  Order 2   &  Order 3    &  Order 4    \\
		& (FCpol1\_d) &(FCpol2\_d) & (FCpol3\_d) & (FCpol4\_d) \\
		\hline	
		$C_0$      & $66.72$  & $65.97$  &  $67.52$  &  $69.41$  \\
		$\sigma_0$ & $0.1654$ & $0.2083$ &  $0.2430$ &  $0.2711$  \\
		$C_1$      & $0.6002$ & $0.6198$ &  $0.5432$ &  $0.3938$ \\
		$\sigma_1$ & $1.253\,10^{-3}$ & $3.572\,10^{-3}$ &  $7.185\,10^{-3}$ &
		$1.204\,10^{-2}$ \\
		$C_2$      &                  & $-7.068\,10^{-5}$ &  $5.561\,10^{-4}$ &
		$2.613\,10^{-3}$ \\
		$\sigma_2$ &                  & $1.207\,10^{-5}$ &  $5.246\,10^{-5}$ & 
		$1.432\,10^{-4}$ \\
		$C_3$      &                  &                  & $-1.260\,10^{-6}$ &  
		$-1.033\,10^{-5}$ \\ 
		$\sigma_3$ &                  &                  &  $1.027\,10^{-7}$ & 
		$5.965\,10^{-7}$ \\
		$C_4$      &                  &                  &                   & 
		$1.225\,10^{-8}$ \\
		$\sigma_4$ &                  &                  &                   &
		$7.937\,10^{-10}$ \\
		\hline	
	\end{tabular}
\end{table}

\begin{table}
	\caption{\small Coefficients of the linear fits to the daily data in the restricted linear range $S_{\mathrm{N}}=5$ - 290 by ordinary least-squares and by orthogonal distance regression. The two fits match closely.}
	\label{T:LinRegd}
	\centering
	\begin{tabular}{lll}
		\hline	
		Coefficients & Order 1          & Order 1 (ODR)    \\ 
		             & (FCRpol1\_d)     &                  \\ 
		$C_0$        & $64.79$          & $62.23$          \\
		$\sigma_0$   & $0.2010$         & $0.2030$         \\
		$C_1$        & $0.6171$         & $0.6419$         \\
		$\sigma_1$   & $1.592\,10^{-3}$ & $1.610\,10^{-3}$ \\
		\hline	
	\end{tabular}
\end{table}

\subsection{Comparison with the Tapping and Morgan (2017) proxy}

\begin{figure}
	\centering
	\subfigure{\includegraphics[width=0.7\columnwidth,trim=0cm 0.2cm 0cm 0.5cm,clip]{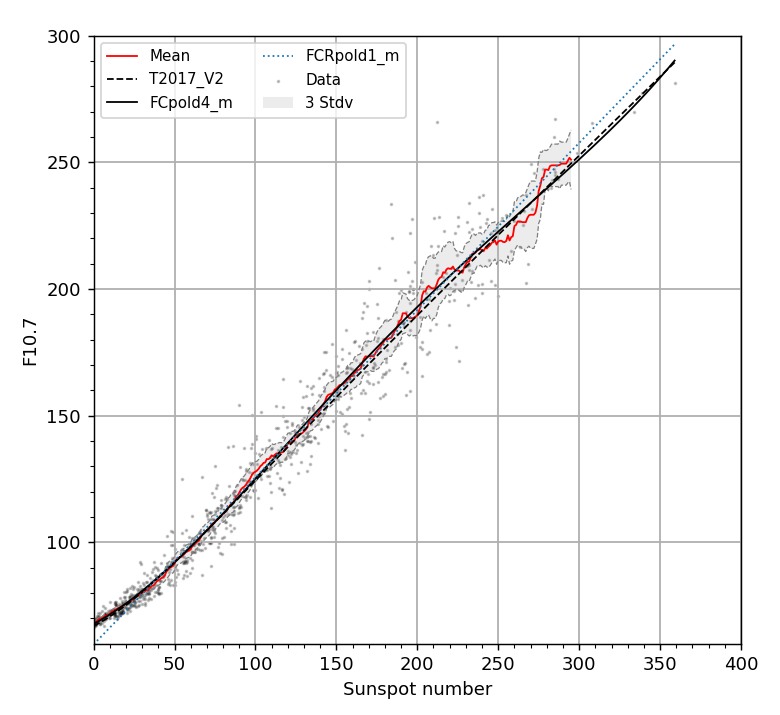}}
	\subfigure{\includegraphics[width=0.7\columnwidth,trim=0cm 0.2cm 0cm 0.5cm,clip]{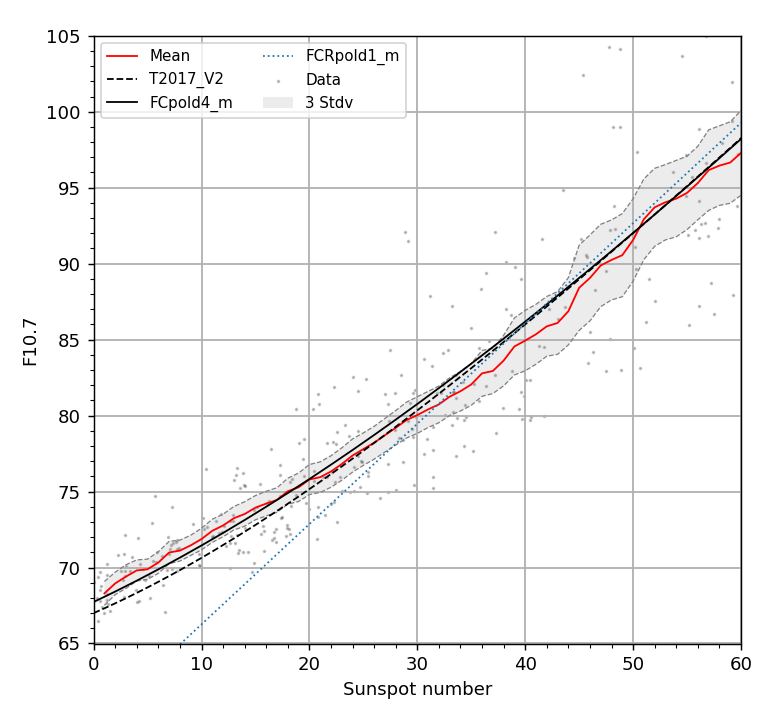}}
		\caption{\small 
			Comparison of our $4^{th}$ order polynomial with the proxy relation from \citet{TappingMorgan2017}. The curves are superimposed on the corresponding non-parametric mean (cf. Fig. \ref{F:MeanCurve_m}) to show the agreement within $3\,\sigma_m$, and on the base data (gray dots). The lower plot is a close-up view of the low range of the upper plot. 	\label{F:Polyn4Compar}} 
\end{figure}

In order to check if indeed the new polynomial fits bring an improvement on past relations, in Figure \ref{F:Polyn4Compar}, we compare the order-4 polynomial with the best curve identified among the past published proxies, namely the curve by \citet*{TappingMorgan2017}.  Here, we consider the fit on monthly means, as most of the fits are based on temporally averaged numbers, in order to smooth out the random variations due to short-timescale solar variations. 

One can see that both fits match very closely, within 4 sfu over the whole linear range.  The slope is slightly lower than the slope of a purely linear fit on the range $S_{\mathrm{N}}= 25$ to 290 (dotted line). This may be due to the influence of the upward deviation for $S_{\mathrm{N}}$ below 30.
 
Now, looking at the lowest range, we find that the $4^{th}$-degree polynomial tracks the data slightly better, and at least remain within the uncertainty range of the means, contrary to the relation by \citet{TappingMorgan2017}, which is too low. It reaches 68 sfu at $S_{\mathrm{N}}= 0$, instead of the 67 sfu background value used by \citet{TappingMorgan2017}. This is in agreement with our analysis of the background flux for a spotless Sun in Section \ref{S:BackgndFlux} below: our polynomial fits the most probable flux instead of the assumed $F_{10.7}$ background value at $S_{\mathrm{N}}= 0$, which is a lower boundary. Finally, we point out that our $4^{th}$-order polynomial is entirely defined by a least-square fit to the data, and is not attached to a predefined tie-point, like the \citet{TappingMorgan2017} curve. It thus allows classical statistical tests on fitted polynomials, including the estimate of errors on polynomial coefficients and on the resulting proxy values. 

\section{Polynomial error determination}

\subsection{Uncertainties on polynomial values}

Regression methods allow to determine the standard errors $\sigma_n$ on each polynomial coefficient. However, an exact derivation of the standard error $\sigma_p$ on polynomial values themselves, based on those standard errors $\sigma_n$, does not exist in the literature, due to the mathematical complexity of this problem. Indeed, the errors on the coefficients for the different terms are actually inter-correlated, as they are determined together. Therefore, the total variance of the polynomial values is not the simple naive sum of the individual variances of all terms. However, a proper estimate of $\sigma_p$ can be derived, based on the fact that the actual error for each term (each degree) is the conditional error on that coefficient, i.e. the uncertainty of that polynomial term given the values of the coefficients for all other terms (in the solution of the least-square regression). In order to estimate this conditional error, we can make a regression for only one term (one polynomial degree) at a time, after subtracting all other terms from the original observed $F_{10.7}$ values, with the other coefficients set at the values given by the regression.

In order to simplify this calculation, we considered that, as we go to higher degree terms, their contribution becomes smaller. Therefore, we derived the conditional error for each degree $n$ by regressing for each degree separately (one-term model), after subtracting successively all polynomial contributions \emph{of lower degrees ($< n$)} from the original $F_{10.7}$ data, starting from the lowest degree. Thus, the single-term model of degree $n$ is: 
\begin{equation}
F_n=  C_n\, S_{\mathrm{N}}^n  \;\;\; n = 0,\dots, d
\label{E:PolyMod}
\end{equation}
where $C_n$ is the coefficient to be determined (with its error) and $d$ is the degree of the polynomial. This model is fitted to the $F_{10.7}$ data series, minus all fitted terms of lower degree:

\begin{equation}
F_{corr}=  F_{10.7} -  \sum_{k=0}^{n-1} C_k\, S_n^k 
\label{E:Fcorr}
\end{equation}

The fact that we did not subtract the terms of higher degrees leads to a slight overestimate of the residual error for each degree, as it also includes the residual uncertainties of all degrees above $n$. Therefore, the conditional errors calculated in this way for each separate degree give an upper limit.

Deriving the above errors from the data requires a statistical processing and multiple regressions on the source data. So, for practical applications, this approach would be too heavy. Therefore, based on the data-based errors obtained by this procedure, we found a simple mathematical representation that gives a good approximation of the $\sigma_p$ errors from the full determination described above, and that can be calculated directly for a polynomial value calculated at any given $S_{\mathrm{N}}$:

\begin{equation}
\sigma_{Tot} = \sqrt{ \sum_{n=0}^{d} \left( \frac{S_N^n - {\overline{S}_{\mathrm{N}}}^n} {2^{\,n-1}\, (d+1-n )^2} \; \sigma_n \right)^2 }
\label{E:PolyErr}
\end{equation}
where $\overline{S}_{\mathrm{N}}$ is the mean of all $S_{\mathrm{N}}$ values in the data set ($\approx 120$ with the actual data).

It consists in the sum of squared errors for each term, using the (non-conditional) standard error on each coefficient given by the least-square regression procedure, $\sigma_n$, but with a weight factor that decreases for increasing degree $n$ (powers of 2), and also decreases for each term of a given degree $n$, as the degree $d$ of the polynomial increases (and thus the number of degrees of freedom in the regression). 

\begin{figure}
	\centering
	\subfigure{\includegraphics[width=0.7\columnwidth,trim=0cm 0.2cm 0cm 0.5cm,clip]{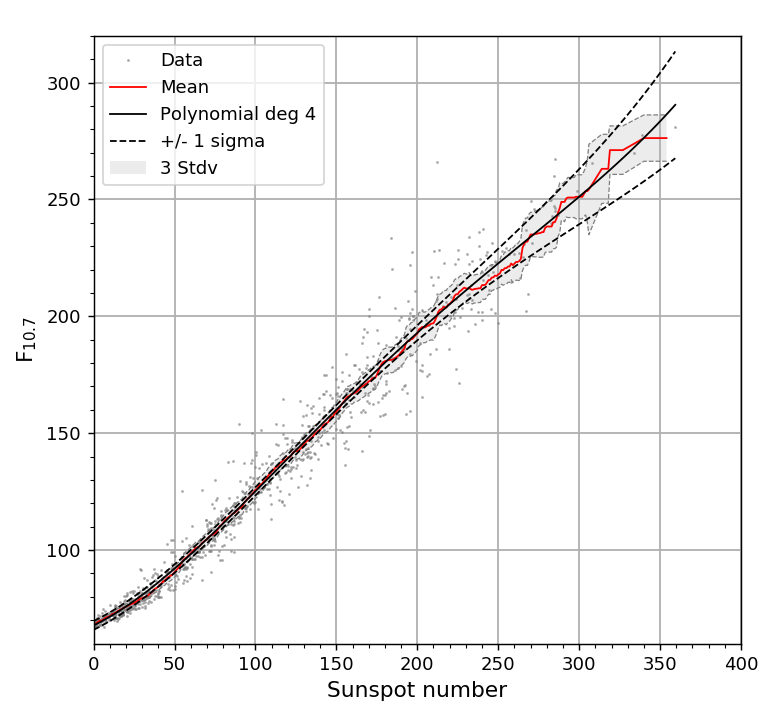}}
	\subfigure{\includegraphics[width=0.7\columnwidth,trim=0cm 0.2cm 0cm 0.5cm,clip]{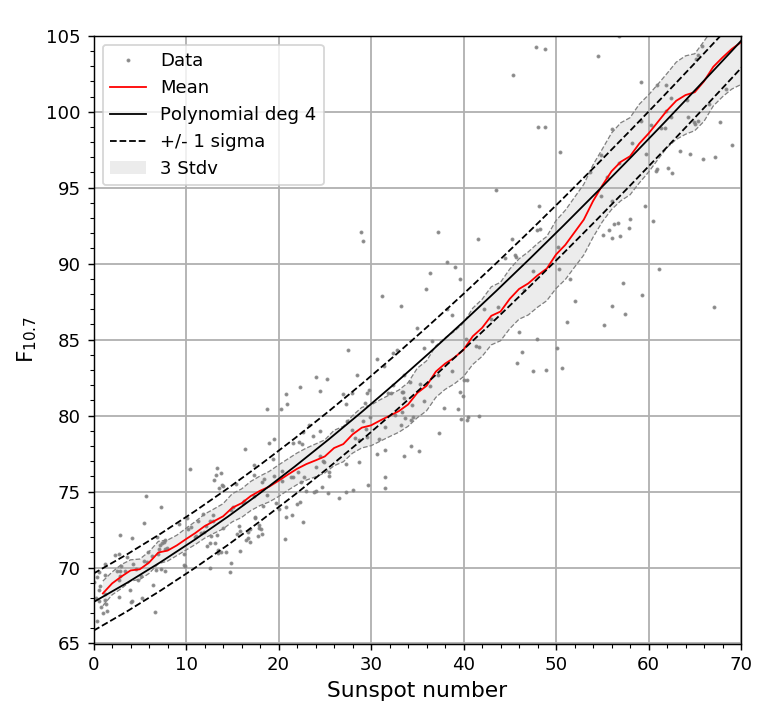}}
		\caption{\small 
			Order-4 polynomial with uncertainty range (1 standard error $\sigma_p$) on polynomial values, superimposed on the corresponding non-parametric mean profile and SEM $\sigma_m$ obtained in Section \ref{S:MeanProf}. The lower plot is a close-up view of the low-activity range in the upper plot. \label{F:Polyn4Err}} 
\end{figure}

We observe that those rather simple expressions already give a very good agreement with the real data-based errors. This rather simple empirical weighting thus probably reflects the dominant corrections associated with the inter-dependency of the least-square polynomial coefficients. A mathematical demonstration goes well beyond the scope of this study, but the good match with the data-based errors indicates that we obtain here a reliable estimate of this polynomial error (Figure \ref{F:Polyn4Err}). This marks a big improvement on all previous proxy relations, where the error was missing and thus entirely undetermined.  The formula in Equation \ref{E:PolyErr} conveniently allows a direct calculation of the error, without requiring to re-do the above extraction of conditional errors from the data themselves. Finally, we point out that $\sigma_{Tot}$ gives the uncertainty on the proxy values, which combines both the errors in $F_{10.7}$ and $S_{\mathrm{N}}$. This must be distinguished from the standard error of a single daily $F_{10.7}$ measurement, which is globally estimated at about 2\% by \citep{TappingCharrois1994}. As expected for such a regression, the smallest errors  ($\pm 2$ sfu) are found in the vicinity of the mean of all values, i.e. $S_{\mathrm{N}}= 120$ and $F_{10.7}=135$ sfu, and the error grows in both directions away from this point.

\subsection{Orthogonal-distance regression versus ordinary least-square regression} \label{SS:PolynODR}

In the ordinary least-square (OLS) regression, the model assumes that all errors are in the dependent variable (“response”, here $F_{10.7}$) and not in the independent variable (“explanatory”, here $S_{\mathrm{N}}$). As we  know that in our case, both quantities are actually affected by errors, we repeated the regression, but using instead the orthogonal distance regression (ODR) technique, which takes into account the uncertainties in both regressed variables. 

\begin{figure}
	\centering
	\subfigure{\includegraphics[width=0.7\columnwidth,trim=0cm 0.2cm 0cm 0.5cm,clip]{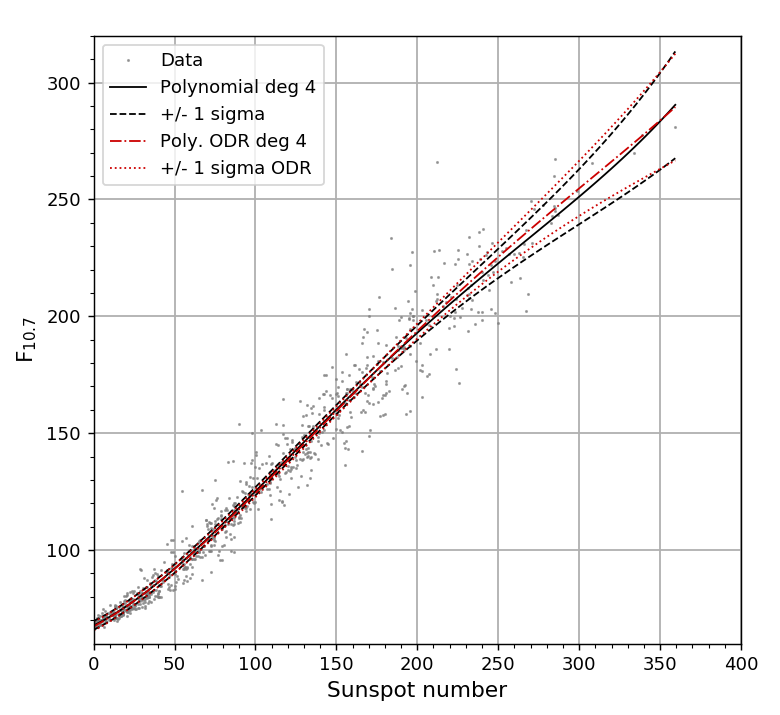}}
	\subfigure{\includegraphics[width=0.7\columnwidth,trim=0cm 0.2cm 0cm 0.5cm,clip]{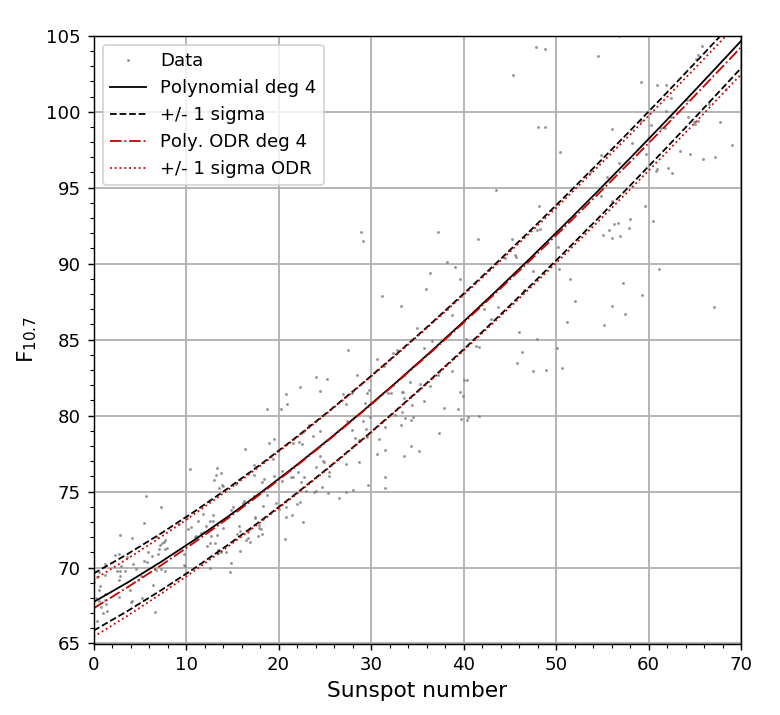}}
		\caption{\small 
			Comparison of $4^{th}$-degree polynomials obtained by the ordinary least-square regression (OLS) and by the orthogonal-distance regression (ODR). The differences are small in comparison with the 1-$\sigma_p$ standard error of the fit, and are thus not significant over the whole range of values.	\label{F:Polyn4_ODR}} 
\end{figure}

We find that the differences between the coefficients derived from the ordinary and ODR regressions are within the computed uncertainties, and are thus not significant.  Likewise, Figure \ref{F:Polyn4_ODR} illustrates this close agreement for the $4^{th}$-degree polynomials derived by both methods. Therefore, we conclude that the ordinary regression gives valid fits in this case. This can be explained by the very high level of correlation between the two indices over long timescales.

\section{Background flux for a spotless Sun} \label{S:BackgndFlux}

In the above curves, we noted that past relations found by \citet{TappingValdes2011} and \citet{TappingMorgan2017} assumed a base radio flux of 67 sfu at $S_{\mathrm{N}}=0$. By contrast, our mean curves based on daily values indicate a higher mean $F_{10.7}$ for all spotless days in the series, at 70.5 sfu. However, the monthly means tend to converge towards lower values near $S_{\mathrm{N}}=0$, though still above 67 sfu. We can thus wonder how to reconcile those apparently contradictory determinations of the same parameter. As the underlying temporal resolution is different in each case, we suspected that the temporal scale plays a central role.

\subsection{Dependency on spotless duration}
\begin{figure}
	\centering
	\includegraphics[width=0.9\columnwidth]{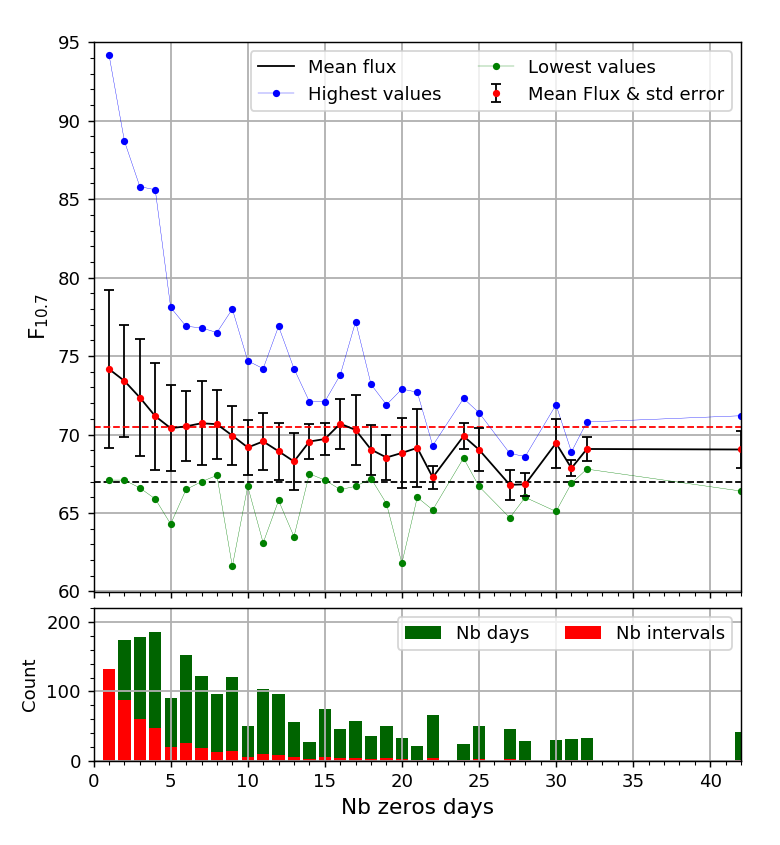}
		\caption{\small 
			Plot of the mean $F_{10.7}$ background flux for spotless days (red dots) as a function of the duration of the sequence of contiguous spotless days.	The errors bars correspond to one standard deviation. The upper curve (blue dots) and lower curve (green dots) are the lowest and largest $F_{10.7}$ values for each spotless duration. The red dashed line marks the overall mean spotless flux (70.5 sfu), while the lower black dashed line is the mean minimum flux (67 sfu). The lower panel gives the number of intervals for each spotless duration (red bar) and the number of days included in each duration (green). \label{F:FluxSpotless}} 
	
\end{figure}

In order to investigate such a temporal effect, we extracted all spotless days in the SN series, and the corresponding daily $F_{10.7}$ flux. Then, we also grouped uninterrupted sequences of contiguous spotless days. Finally, we computed the distribution of $F_{10.7}$ values for all spotless sequences of the same length in the observed series. Figure \ref{F:FluxSpotless} shows the mean values, standard deviations and extreme values of the $\mathrm F_{10.7cm}$ flux for all sequence lengths found in the series. In the lower panel, we also plotted the number of days included in each category, and how many sequences were found for each duration. The longest sequence lasted 42 days, but most spotless days sequences last less than 10 days, with many isolated spotless days.

Our analysis shows that the mean $F_{10.7}$ flux systematically increases as the duration of a spotless-day sequence decreases. For duration above 15 days, the most likely $F_{10.7}$ value is near 68 or 69 sfu. On the other hand, it increases to 74 sfu for single spotless days immediately surrounded by active days with one or more sunspots. The lowest mean daily value is 67 sfu and is reached only for 6 sequences with duration of 22, 27 and 28 days. The red dashed line at 70.5 sfu corresponds to the mean flux for all spotless days. Quite logically, it corresponds to the mean levels for duration 5 to 10 days, which is near the mean duration of spotless intervals.

Now, considering the extreme values, we also find a steady increase of the upper values (blue dots) with decreasing duration, up to as high as 95 sfu for a single spotless day. Based on the above regressions, such fluxes usually correspond to $S_{\mathrm{N}}$ values of 50, i.e. to moderate levels of activity. On the other hand, the lowest values (green dots) do not show any dependency on duration. They stay around the 67 sfu level, with rare extremely low values down to 61.6 sfu (November 3, 1954).  This thus validates the  choice of 67 sfu as the all-quiet base flux, shown in Figure \ref{F:FluxSpotless} as the horizontal black dashed line. 

Actually, there are only 33 values below 66 sfu in the whole series. Moreover, all 8 values below 65 sfu and 14 values out of 25 between 65 and 66 sfu appear exclusively in 1953 and 1954, some of them even on days when the Sun was not spotless. By contrast, the lowest values recorded after 1954 are always above 65.5 sfu, and almost all are occurring quite logically during the longest minimum recorded in the $F_{10.7}$ series, between cycle 23 -- 24 in 2008. By comparison, the cycle 18 -- 19 minimum in 1954 was not particularly low and protracted. This suggests that the record-low values in 1953 -- 1954 are either spurious or suffer from a calibration problem. As indicated by \citet{Tapping2013} and \citet{TappingMorgan2017}, those early data may indeed suffer from larger errors, as the calibration method was not fully standardized, and the location in Ottawa until 1962 caused larger radio interferences. Therefore, a base background flux as low as 64 sfu, as suggested by \citet{TappingDetracey1990} and \citet{TappingCharrois1994}, seems doubtful and too low for the real fully-quiet Sun.

\subsection{Interpretation}

We can explain this dependency of the background levels on the duration of spotless intervals by the presence of other sources of $F_{10.7}$ emission, even when there are no associated sunspots. This includes various features in the chromosphere and lower corona associated with closed magnetic fields weaker than those concentrated in sunspots, like  bright chromospheric plages, filaments, coronal condensations \citep{ShimojoEtal2006,SchonfeldEtal2015,PevtsovEtal2014,ErmolliEtal2014}. In particular, when the Sun is very quiet (few isolated spots) or entirely spotless, the associated plages have a small extent but still contribute a significant excess in $F_{10.7}$. Indeed, Figure \ref{F:FluxSpotless} shows that most of the values are below 80 sfu, although there are a few as high as 80 sfu or more. On the other hand, there are almost no values below 67 sfu. 

Such chromospheric plages are typically present just before the emergence of a first spot, or they remain after the decay of a last sunspot.  Therefore, short spotless intervals surrounded by more active periods are most likely to show a significant excess in radio emission above the lowest all-quiet level. Conversely, only very long spotless periods can include days without any activity features on the Earth-facing solar disk. The only small active regions present just before the long spotless interval have enough time to decay entirely, well before new bright chromospheric structures develop, heralding the appearance of the first spots marking the end of the protracted spotless period. 

Consequently, the mean background level does not have an absolute value, as the lowest level of 67 sfu is almost never reached, even when there are no sunspots. We can thus expect to see a dependency of the asymptotic $F_{10.7}$ value near $S_{\mathrm{N}}=0$ when applying a temporal averaging to the raw daily data series.  Without any averaging or when the averaging duration is short, days in long spotless intervals are mixed with those in short intervals, leading to a higher mean, rising to 70.5 sfu for 1-day timescale. At the other extreme, averaging over very long duration, like one year, inevitably mixes spotless periods with active periods, as virtually no spotless periods have duration above about 30 days. Therefore, the lowest $F_{10.7}$ yearly mean values are expected to be also higher than the 67 sfu minimum. \emph{It turns out that a duration of one month best matches the actual duration range of spotless episodes. Monthly means are thus best for recording the lowest possible mean radio fluxes of the fully quiet Sun.}

Finally, we note that this chromospheric background interpretation agrees with the identification of two types of emission sources for $F_{10.7}$ \citep{Tapping1987,TappingDetracey1990,TappingZwaan2001,TappingMorton2013,SchonfeldEtal2015}. While the gyroresonance emission is closely associated with strong magnetic fields in sunspots ($> 300$ G), a so-called ''diffuse'' component by free-free thermal emission is attributed to plages and the overall chromospheric network. The latter is the best candidate for the variable background flux diagnosed here. In this respect, we note that while the sunspot component of $F_{10.7}$ will track instantaneously the evolution of active regions (fully linear relation), the plage component will be extended and delayed in time relative to the associated sunspots, as it corresponds to the progressive decay and dispersal of flux emerging in active regions. 

Therefore, the dependency between the mean background flux and the duration of the spotless interval is consistent with this interpretation, and offers an independent indicator of this dual-source nature of the $F_{10.7cm}$ radio flux. It also implies that \emph{the disagreements between $F_{10.7}$ and the SN are probably due for a significant part to the time delay intrinsic to the free-free emission from plages}, rather than simply to a non-proportionality with the underlying emerging magnetic fluxes, and thus with the SN. Indeed, the latter is only sensitive to strong fluxes freshly emerged in sunspots, without mixing with a second magnetic-decay component. The fact that disagreements between those two indices increase for short time scales, below one solar rotation and thus below the average plage lifetime, also concurs with the prominent role of this temporally-smeared weak-field component.  

\section{A simple model for the the $\bf F_{10.7}/S_{\mathrm{N}}$ non-linearity} \label{S:DataModel}

In the above analyses, we found that daily values indicate that the relation between $F_{10.7}$ and $S_{\mathrm{N}}$ is linear almost down to the lowest values of $S_{\mathrm{N}}=11$ (single isolated spot). Only for smaller $S_{\mathrm{N}}$ values close to 0, $F_{10.7}$ stops decreasing and reaches its background level, in the range 68-70 sfu as found in Section \ref{S:BackgndFlux}. There is a break from the linear relation, with the last points near $S_{\mathrm{N}}=0$ located several solar flux units above the linear relation, which intercepts the axis at $S_{\mathrm{N}}=0$ at a value of 58 to 62 sfu (see tables of polynomial coefficients above).
  
When deriving monthly or yearly means, the averaging interval inevitably includes periods of different activity levels, including some inactive days, when $F_{10.7}$ is at its lowest level and thus shows an excess relative to a purely linear relation. As the mean activity during the averaging interval decreases, the proportion of  spotless days increases, and thus also the fraction of points bringing an excess above the linear relation. We can thus expect a progressive upwards deviation from linearity, like we observe in the monthly and yearly mean curves.

In order to simulate this scheme, we took the observed SN time series, and synthesized a $F_{10.7}$ time series, by converting each $S_{\mathrm{N}}$ value via a two-component model:
\begin{enumerate}
	\item For most of of the range, we used the linear fit to the linear part of the data (from Table \ref{T:LinRegd}). 
	\item For the lowest $S_{\mathrm{N}}$ values, when the linear relation falls below the base $F_{10.7}$ background, chosen at 70.5 sfu, the output value is set at the constant value of 70.5 sfu. An alternate model for this background flux can take into account the fact that the mean background is not constant but increases even when the Sun is spotless, up to 74 sfu, as demonstrated in Section \ref{S:BackgndFlux}.
\end{enumerate}
We then applied the usual monthly and yearly averaging to this synthetic series. 

\begin{figure}
	\centering
	\includegraphics[width=1.\columnwidth]{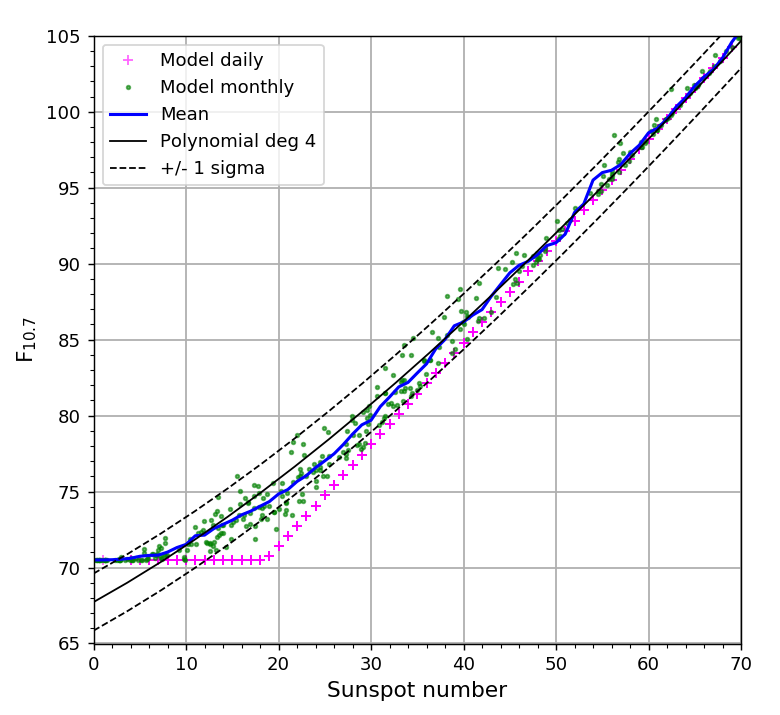}
		\caption{\small 
			Model of the monthly temporal averaging of daily data built from two components: linear component and constant lower background at 70.5 sfu. The pink crosses are the synthesized daily values. The green dots are the corresponding monthly mean values, and the blue line is the non-parametric local mean of those values. As a comparison, the black line is the $4^{th}$-order polynomial fitted to the real monthly mean data (Table \ref{T:PolyOLSm}), with uncertainties (black dashes lines).
				\label{F:Model70f_m}} 
\end{figure}

\begin{figure}
	\centering
	\includegraphics[width=1.\columnwidth]{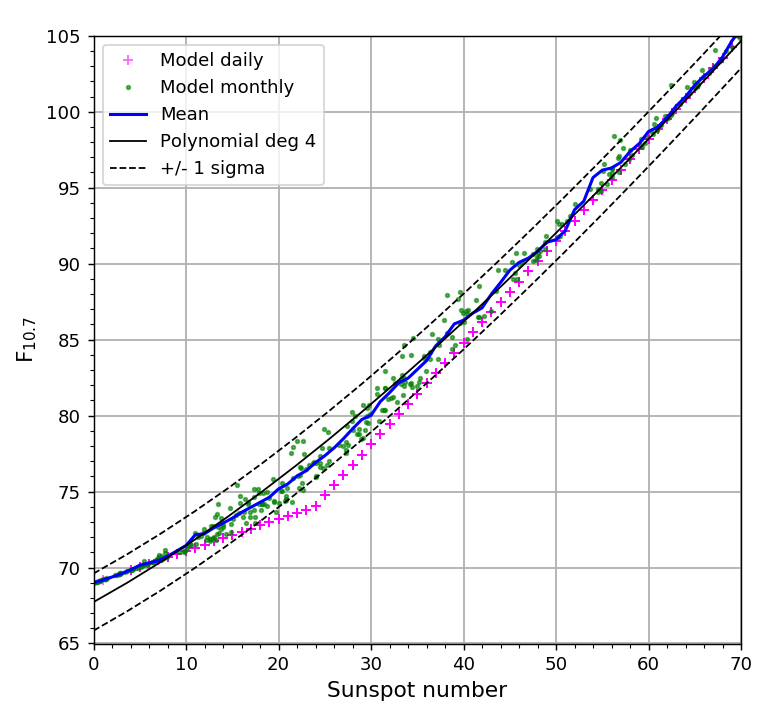}
		\caption{\small 
		 Model of the monthly temporal averaging of daily data built from two components: linear component, and here, a slightly rising background with increasing $S_{\mathrm{N}}$. The elements of the plots are the same as in Figure \ref{F:Model70f_m}.	
			\label{F:Model69t_m}} 
\end{figure}

In Figure \ref{F:Model70f_m}, the model for daily values assumes a constant background at the mean   $F_{10.7}$ flux for a spotless Sun (pink crosses). In Figure \ref{F:Model69t_m}, the model assumes a progressive rise of the $F_{10.7}$ flux over the range found in our above analysis of the base flux according to the duration of the spotless period. It starts from 69 sfu, the low value for a fully spotless Sun and rises to 75 sfu, the upper value found for isolated spotless days. This can be considered as representative of the flux when just one isolated sunspot group is present on the Sun. Here, this ramp connects with the main linear relation at about $S_{\mathrm{N}}=24$. 

Our degree-4 polynomial based on the true data (black line, with uncertainties as dashed lines) nicely falls in the middle of the simulated monthly means for both options. The agreement with the mean of data values (blue curve) is best for the second model, where the agreement is very tight. The first model with a higher but uniform background gives a higher curvature below $S_{\mathrm{N}}=24$.
  
\begin{figure}
	\centering
	\includegraphics[width=1.\columnwidth]{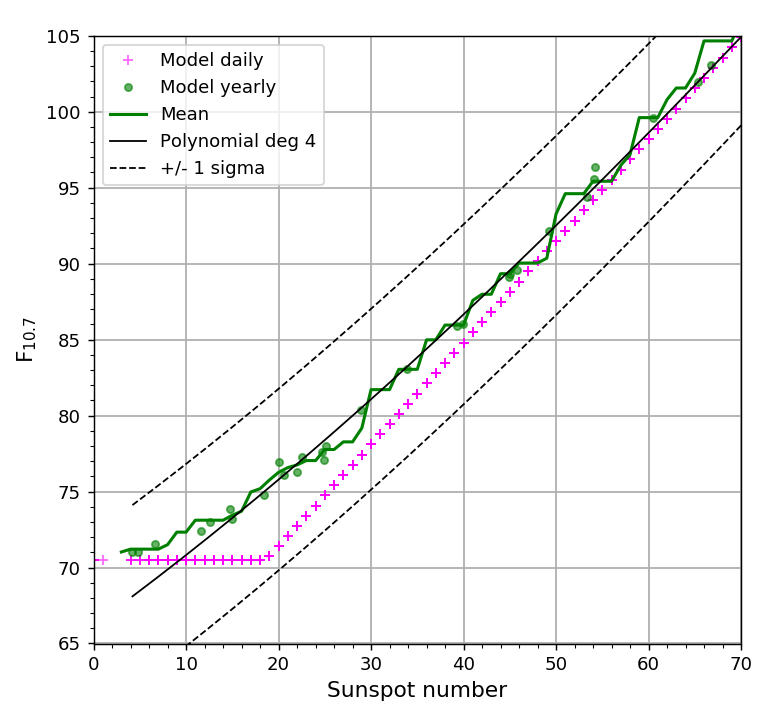}
		\caption{\small 
			Model of the yearly temporal averaging of daily data built from two components: linear relation and constant lower background at 70.5 sfu. The elements of the plots are the same as in Figure \ref{F:Model70f_m}, with green dots and curve corresponding to yearly averages and the polynomial also to yearly mean data (Table \ref{T:PolyOLSy}).
			\label{F:Model70f_y}} 
\end{figure}

\begin{figure}
	\centering
	\includegraphics[width=1.\columnwidth]{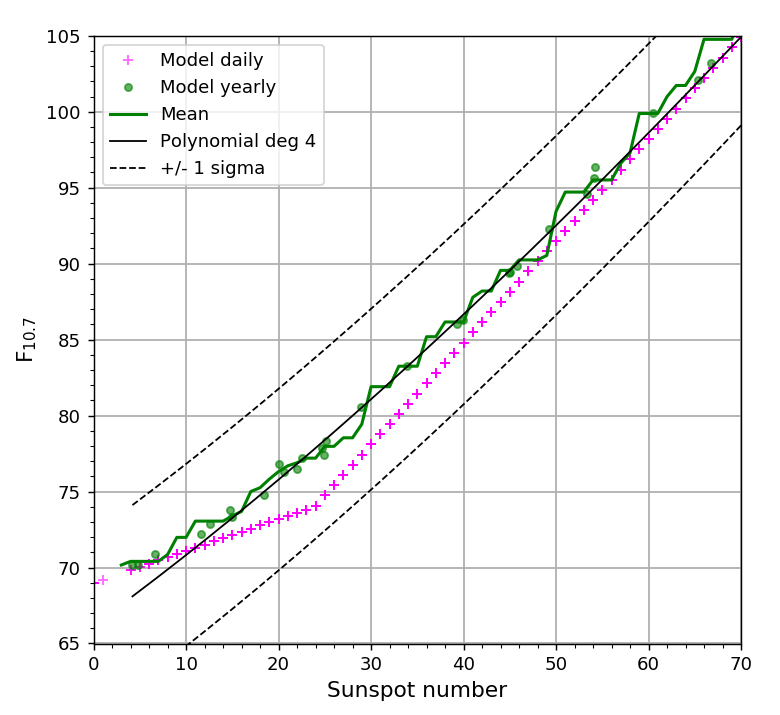}
		\caption{\small 
			Model of the yearly temporal averaging of daily data built from two components: linear relation and slightly rising background with increasing $S_{\mathrm{N}}$. The elements of the plots are the same as in Figure \ref{F:Model70f_m}.	
			\label{F:Model68t_y}} 
\end{figure}

We also made the simulations using yearly means (Fig. \ref{F:Model70f_y} and \ref{F:Model68t_y}). They also give a good agreement, but given the lower number of points and slightly more linear relation, the monthly simulations shown here illustrate more clearly how temporal averaging is producing a curvature of the relation.

Overall, those two very simple simulations match strikingly well the actual data. They thus confirm the mechanism by which time-averaging of the raw linear daily values can produce the non-linear proxy relation, thus also indicating that this non-linearity is dependent on the temporal-averaging applied to the data before making the regression.

\section{Temporal variations} \label{S:TempVar}

So far, we included the entire duration of the time series, thus making the assumption that both the $F_{10.7cm}$ flux and the sunspot number series are homogeneous over the entire 68-year duration included here. However, \citet{CletteEtal2016} made a first simple comparison between the newly released Version 2 of the sunspot number and $F_{10.7}$ as a function of time, and found a 12\% upward jump in the $F_{10.7}/S_{\mathrm{N}}$ ratio, occurring between 1979 and 1983. 

Likewise, by a comparison to the sunspot number series (versions 1 and 2) and the total sunspot area, \citet{TappingValdes2011} and \citet{TappingMorgan2017} found that the $F_{10.7}$ time series shows an upward deviation in the second half of the series, mostly after 1980, relative to both the sunspot number and sunspot areas. Although this trend is stronger when comparing with $S_{\mathrm{Nv1}}$, it is still present when $S_{\mathrm{Nv2}}$ is taken as reference. The authors fit a smooth curve as a function of time over the whole duration of the series, and they interpret the resulting global trend as a real change in solar properties. This evolution would parallel the overall decline of solar cycle amplitudes since the mid-$20^{th}$ century, thus invoking a possible genuine change in the properties of the Sun.

\subsection{A transition between two stable periods}

In order to check for such a change in the relation between the two measurements, we used the 13-month smoothed monthly means, using the classical Z\"urich smoothing function. This allows to reject random fluctuations associated with solar activity at time scales shorted than one year, while retaining a better temporal sampling than in yearly means.  In our case, in the above $F_{10.7}$ versus $S_{\mathrm{N}}$ representation, we checked over which time interval the data follow the same relation in the $F_{10.7}/S_{\mathrm{N}}$ space.
 
\begin{figure}
	\centering
	\includegraphics[width=1.\columnwidth]{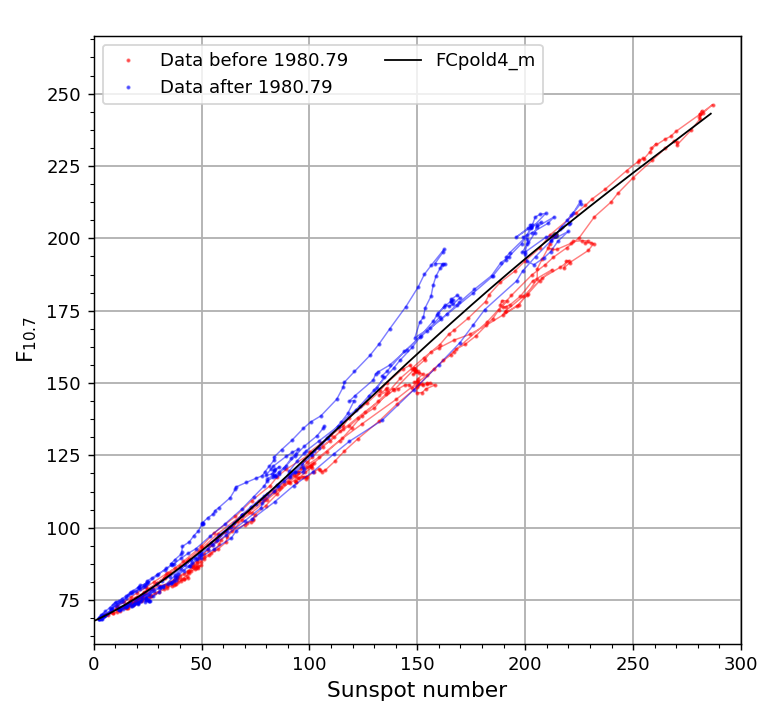}
		\caption{\small 
		Plot of $F_{10.7}$ versus $S_{\mathrm{N}}$, using the original SN series (Version 1). The data are smoothed by a 13-month running mean. The line connects successive months, and thus illustrate the chronological evolution. The curve is colored in blue or red for dates before and after 1981.		
			\label{F:TwoPer_SNv1}} 
\end{figure}

\begin{figure}
	\centering
	\includegraphics[width=1.\columnwidth]{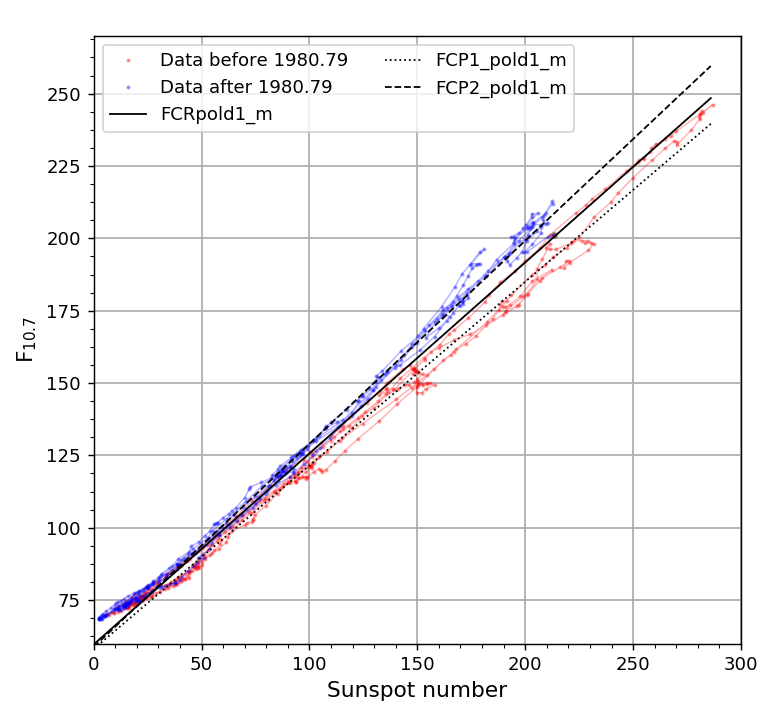}
		\caption{\small 
		Plot of $F_{10.7}$ versus $S_{\mathrm{N}}$, equivalent to figure \ref{F:TwoPer_SNv1} but using the new re-calibrated version of the SN series (Version 2). The data are smoothed by a 13-month running mean. The data curve is in blue or red for dates before and after 1981. The black lines correspond to linear fits to the entire series (solid line), the period before 1980 (dotted) and the period after 1980 (dashed).	
			\label{F:TwoPer_SNv2}} 
\end{figure}

In Figures \ref{F:TwoPer_SNv1} and \ref{F:TwoPer_SNv2}, the resulting curves are shown respectively for the SN version 1 and SN version 2. The curves now include the chronology and consist of several narrow loops corresponding to each of the solar cycles, varying from very low values at minima to the maxima while staying close to a diagonal line, as can be expected given the close proportionality of the two indices, as shown in the previous sections.

We note that two cycles deviate strongly in the case of the SN version 1 (blue loops in Figure \ref{F:TwoPer_SNv1}). They correspond to cycles 22 and 23, when SN version 1 is know to be affected to drifts in the Locarno pilot station \citep{{CletteEtal2014}, {CletteEtal2016}}. On the other hand, the agreement is much tighter with SN version 2 (Figure \ref{F:TwoPer_SNv2}), in particular over all cycles after cycle 21 for which the SN was entirely re-constructed based on a multi-station reference. 

We will thus now concentrate on this second comparison with SN version 2. The plot immediately shows that the curves are grouped along two preferential bands located above (colored blue) and below (red) the fit to the entire series (solid black line), with very few points in-between, near the global fit. This indicates that the relation was actually very stable during two time periods and jumped directly from one relation to the other.

We then looked for the time sub-intervals following the upper and lower linear relation, and we found that the lower relation applies to all data before 1980, while the upper relation is valid for the entire period after 1980. \emph{The temporal evolution is thus characterized by a single jump separating two fully homogeneous periods, during which the $F_{10.7}/S_{\mathrm{N}}$ relation is very stable}. 

For each homogeneous interval, we could then derive the two corresponding linear relations using the same regression methods, as applied before to the whole series (Fig. \ref{F:TwoPer_SNv2}).  For the period before 1980, we find a slope of $0.635 \pm  6.61 10^{-3}$ (dotted line), while after 1980, the slope becomes steeper at $0.702 \pm  8.87 10^{-3}$ (dashed line). This corresponds to an upward jump by a factor $1.106 \pm 0.017$, thus of about 10.5\%. The slope found for the entire series naturally falls in-between, with an intermediate slope of $0.660 \pm  6.313 10^{-3}$ (cf. Table \ref{T:LinRegm}). The coefficients for those two linear fits are given together with the two $4^{th}$-degree polynomial fits in  Table \ref{T:TwoPoly}. Such a jump is highly significant, as \citet{TappingCharrois1994} and \citet{Tapping2013} give an accuracy of 1\% for the flux measurements and of 2\% for the daily index derived from the ''spot'' measurements at 20h00 UT.
\begin{table}
	\caption{\small Coefficients of order-1 (linear) and order-4 polynomials fitted by the ordinary least-square regression on the monthly mean SN for the periods 1947-1980 and 1981-2015, with standard errors $\sigma_n$ for each coefficient.}
	\label{T:TwoPoly}
	\centering
	\begin{tabular}{lllll}
\hline		
  Coefficients & Period 1947-1980 & Period 1981-2015 & Period 1947-1980 &
  Period 1981-2015 \\
             &  Order 1 & Order 1  & Order 4  &  Order 4    \\
  \hline
  $C_0$      & $58.09$  & $58.78$  & $66.64$  & $67.85$  \\
  $\sigma_0$ & $0.9391$ & $1.151$  & $1.476$  & $1.275$  \\
  $C_1$      & $0.6345$ & $0.7020$ & $0.3658$ & $0.3845$ \\
  $\sigma_1$ & $6.612\,10^{-3}$ & $8.867\,10^{-3}$ & $6.765\,10^{-2}$ &
   $8.115\,10^{-2}$ \\
  $C_2$      &   &   &  $2.587\,10^{-3}$ &      $2.881\,10^{-3}$  \\
  $\sigma_2$ &   &   &  $8.642\,10^{-4}$ &      $1.378\,10^{-3}$  \\
  $C_3$      &   &   & $-9.906\,10^{-6}$ & $\it -7.429\,10^{-6}$  \\
  $\sigma_3$ &   &   &  $4.029\,10^{-6}$ & $\it  8.344\,10^{-6}$  \\
  $C_4$      &   &   &  $1.329\,10^{-8}$ & $\it  2.694\,10^{-10}$ \\
  $\sigma_4$ &   &   &  $6.152\,10^{-9}$ & $\it  1.645\,10^{-8}$  \\
\hline
	\end{tabular}
\end{table}

Checking the past published proxies, we note that early proxies that were based only on the first part of the series, like \citet{HollandVaughn1984} or the IPS formula (Table \ref{T:ListProxies}), match the low linear fit for the period before 1980 in Table \ref{T:TwoPoly}, while the NOAA proxy adjusted to the recent SN version 2 (\ref{T:ListProxies}), has a steeper slope matching well the second higher fit for the recent period after 1980. So, the disagreements between some of the past proxies actually originate from the temporal inhomogeneity of the $F_{10.7}$ index itself, as diagnosed here.

Another consequence is that when including the entire time series, the least-square regression errors will not decrease when using monthly or yearly mean values, exactly as we found in our global regressions (sub-Sections \ref{SS:PolynMonth} and \ref{SS:PolynYear}). We can now explain it by the fact that a significant part of the deviations of individual monthly or yearly mean  values relative to the mean regression curve is due to this systematic inhomogeneity in the series, rather than to random errors, and will thus not be reduced by temporal averaging.

\subsection{A more precise timing of the jump}

Using the 13-month smoothed monthly values, we can roughly locate the jump in mid-1980, as the smoothed monthly means migrate from the low branch to the high branch between January 1980 and March 1981. However, with such smoothed values, we cannot pinpoint the transition time with a better temporal resolution. Still, the fact that the branch-to-branch transition occurs over a duration similar to the smoothing duration already indicates that the actual transition must take place over a duration much shorter than 13 months, thus very sharply. 

\begin{figure}
	\centering
	\includegraphics[width=1.\columnwidth]{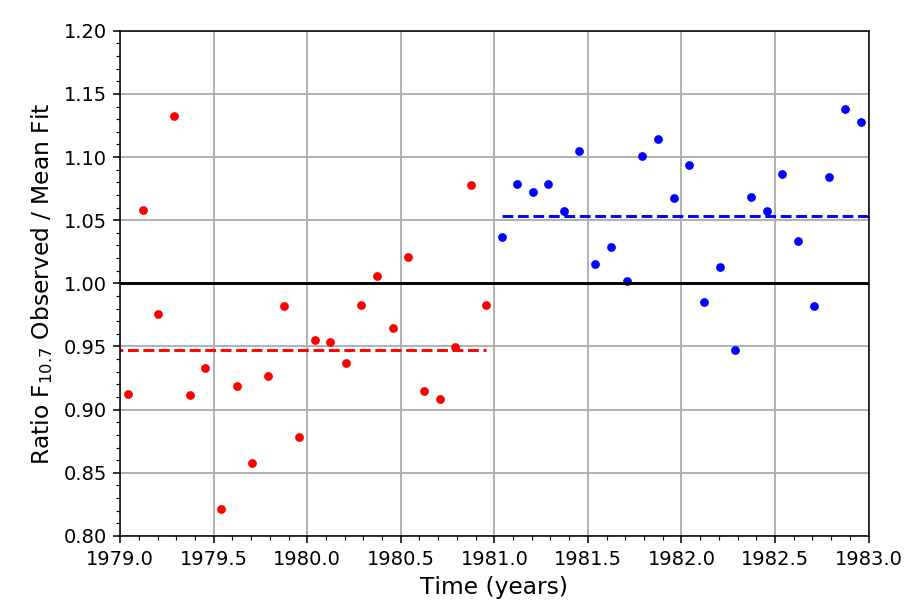}
		\caption{\small 
			Monthly ratios between the observed monthly mean $F_{10.7}$ flux and the proxy value derived from the linear fit over the entire interval 1947 -- 2015  (Table \ref{T:LinRegm}).	The dashed lines correspond to the global fits on the entire half-series before and after 1980.
			\label{F:F107jump}} 
\end{figure}

In order to better pinpoint the time of the transition, we checked the (un-smoothed) monthly means, and we compared them with the global linear fit to the whole data set (Table \ref{T:LinRegm}). In Figure \ref{F:F107jump}, we plot the resulting monthly ''observed/fit'' ratios over the 4-year interval around the suspected transition, which is also centered on the maximum of cycle 21. Over this interval, all data values are thus in the same high range, and the choice of mean fit has only a small influence on the ratios. As can be expected, the monthly ratios show rather large month-to-month random fluctuations, but 19 out of 24 ratios are below 1 before the end of 1980, while 21 out of 24 are above 1 after 1980, indicating a clear systematic change. The points are actually grouped respectively around the global linear fits calculated for the complete half-series 1947 -- 1980 and 1980 -- 2015 (dashed lines). 

Moreover, this transition happens very sharply. With the exception of November 1980, there is a sudden jump between December 1980 and January 1981, followed by 8 consecutive months in 1981 all above the mean linear fit. \emph{This thus strongly suggests that the jump occurred between December 1980 and January 1981, i.e. at the transition between two calendar years}. Although a slightly earlier transition is not entirely excluded, it cannot be before mid-1980.

\subsection{Validation against multiple independent stations}

Now, it turns out that 1980--1981 also marks a transition for the sunspot number series. This is when the production of the sunspot number moved from Zurich to Brussels, with a significant change in the production method \citep{{CletteEtal2014},{CletteEtal2016}}. The former manual processing was computerized and the Specola Solare Observatory in Locarno (Z\"urich's auxiliary station) took over as pilot station in replacement for the Z\"urich Observatory. Therefore, as the sunspot number is a synthetic index based on a global statistical processing of multiple data sources, we should be careful that the 1980 jump that we just found above is not entirely due to a sharp change of scale between the Z\"urich and Brussels sunspot numbers. 

\begin{figure}[t]
	\centering
	\includegraphics[width=1.\columnwidth]{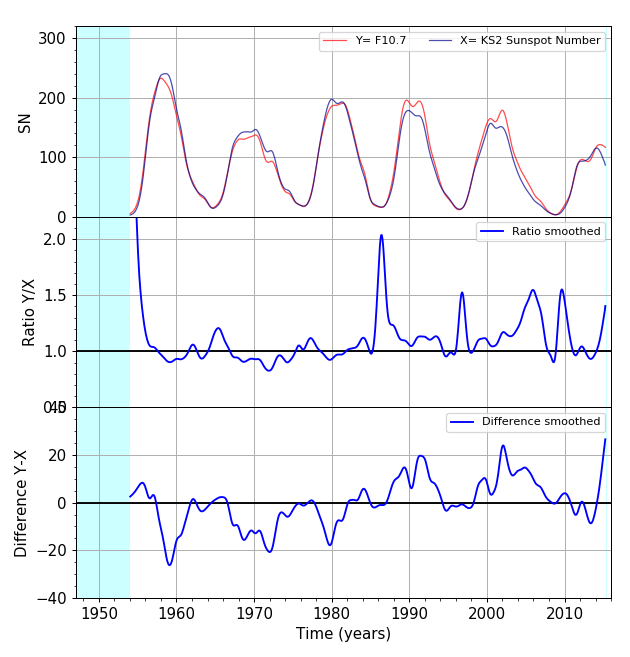}
		\caption{\small 
		Comparison of the $F_{10.7}$-based SN proxy series with the raw Wolf numbers from the Kislovodsk solar Observatory (Russia; SILSO station ID: KS2), cover the period before and after 1980. The upper panel shows the two series, smoothed by a 12-month Gaussian kernel to reduce short-term random noise. The middle and lower panels show the ratios and differences. No data are available for that station in the blue-shaded time interval. An upward jump can been seen around 1980.	
			\label{F:F107vsKS2}} 
\end{figure}
 
\begin{figure}
	\centering
	\includegraphics[width=1.\columnwidth]{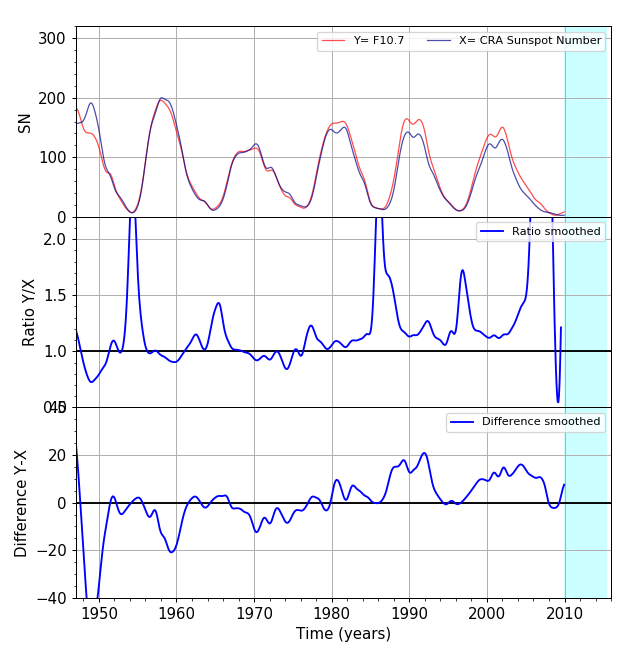}
		\caption{\small 
		Comparison of the $F_{10.7}$-based SN proxy series with the raw Wolf numbers from Thomas Cragg (SILSO station ID: CRA), an amateur sunspot observer, who was employee of the Mount Wilson Observatory. Like in Figure \ref{F:F107vsKS2}, an upward jump in the ratio and differences (middle and lower panels) can been seen around 1980.		
			\label{F:F107vsCRA}} 
\end{figure}

In order to exclude any processing flaw in the sunspot number, we made direct comparisons with a large set of individual stations that provided sunspot counts over a long period extending both before and after 1980. In the SILSO database, we found 28 stations fulfilling the requirements. In fact, most of those stations were part of the multi-station reference used for the re-calibration of the SN version 2 \citep[Table 1]{CletteEtal2016}.  For each station, we derived the ratio between the raw Wolf numbers and the corresponding $F_{10.7}$-based SN proxy value, as a function of time (single proxy relation for all times). As only one station is used in each comparison, this ratio is affected by a larger error. Therefore, we averaged the ratios for the whole time intervals before and after 1980, and then looked at the ratio between the two intervals. This is listed in Table \ref{T:StationRatios}. As illustration, Figures \ref{F:F107vsKS2} and \ref{F:F107vsCRA} show the comparisons for two sample stations: the Kislovodsk station (Observatory of Pulkovo), a professional observatory, and Thomas Cragg, a dedicated individual observer, who was employee of the Mount Wilson Observatory \citep{HoweClette2015}.

\begin{table}
	\caption{\small k ratios between the raw Wolf number from 28 stations in the SN database (World Data Center SILSO) and a fixed $F_{10.7}$-based proxy of the sunspot number, for observations made before 1980 (column 3) and after 1980 (column 5). For each station, we also list the corresponding time interval over which the data are available (columns 2 and 4).  The absolute value of those ratios is different for each station, due to the different observing setup (telescope, observing site), but  it is not relevant here, as we only trace relative differences between different times. The $6^{th}$ column gives the $k_2/k_1$ ratios between the scale ratios for the two periods. Stations marked with a * in the first column form a subset of stations with the best reliability and/or longest duration before and after the 1980 transition.}
	\label{T:StationRatios}
	\centering
	\begin{tabular}{llllllll}
  \hline
  Station & Period 1 & $k_1$ & Period 2 & $k_2$ & Ratio & Error & Comment \\
  \hline
  AN *   & 1976-1980 & 1.435 & 1981-1988 & 1.608 & 1.121 & 0.030 & short series            \\
  AT     & 1968-1980 & 0.878 & 1981-1982 & 1.096 & 1.248 & 0.050 & short series            \\
  BN-S   & 1965-1980 & 0.995 & 1981-2014 & 0.870 & 0.874 & 0.016 & decreasing              \\
  BRm    & 1974-1980 & 0.819 & 1981-1998 & 1.050 & 1.282 & 0.015 & short series            \\
  CA     & 1949-1980 & 1.174 & 1981-2015 & 1.097 & 0.934 & 0.013 & drifts before 1980      \\
  CRA *  & 1947-1980 & 1.267 & 1981-2010 & 1.521 & 1.200 & 0.011 & long series             \\
  EB     & 1949-1981 & 0.746 & 1981-2015 & 1.291 & 1.731 & 0.012 & drifts before 1980      \\
  FR-S   & 1976-1980 & 0.790 & 1981-1988 & 0.962 & 1.218 & 0.019 & short series            \\
  FU  *  & 1968-1980 & 1.047 & 1981-2015 & 1.123 & 1.072 & 0.012 & long and stable         \\
  GU-S   & 1974-1980 & 1.617 & 1981-1991 & 1.936 & 1.197 & 0.053 & short series            \\
  HD-S   & 1967-1980 & 1.151 & 1981-2013 & 1.315 & 1.142 & 0.024 & unstable                \\
  HU     & 1969-1980 & 1.323 & 1980-2015 & 1.164 & 0.880 & 0.024 & decreasing, unstable    \\
  KH  *  & 1966-1980 & 1.098 & 1980-2015 & 1.343 & 1.223 & 0.015 & long series             \\
  KOm    & 1947-1980 & 1.016 & 1981-1996 & 1.198 & 1.178 & 0.016 & slight drift after 1983 \\
  KS2 *  & 1954-1980 & 1.069 & 1981-2015 & 1.219 & 1.140 & 0.009 & long series             \\
  KZm *  & 1947-1980 & 1.026 & 1981-2015 & 1.074 & 1.047 & 0.011 & long series             \\
  LFm *  & 1947-1980 & 1.136 & 1981-1988 & 1.403 & 1.235 & 0.023 &                         \\
  LK  *  & 1967-1980 & 1.321 & 1981-1987 & 1.457 & 1.103 & 0.022 & short series            \\
  LO     & 1958-1980 & 0.822 & 1981-2015 & 0.945 & 1.150 & 0.007 & drifts after 1983       \\
  MA  *  & 1971-1980 & 1.039 & 1981-1988 & 1.183 & 1.139 & 0.015 & short series            \\
  MD     & 1978-1980 & 0.988 & 1981-1986 & 1.532 & 1.551 & 0.068 & short series            \\
  PO     & 1950-1980 & 1.122 & 1981-2000 & 1.106 & 0.986 & 0.017 & constant ratio          \\
  QU     & 1957-1980 & 1.411 & 1981-2015 & 1.820 & 1.290 & 0.019 & unstable                \\
  SA     & 1957-1980 & 1.291 & 1981-2000 & 1.692 & 1.311 & 0.026 & long series             \\
  SC-S   & 1960-1980 & 1.396 & 1981-2007 & 1.133 & 0.812 & 0.021 & decreasing              \\ 
  SK     & 1950-1980 & 1.100 & 1981-2012 & 1.281 & 1.165 & 0.015 & unstable                \\
  SM     & 1967-1980 & 0.794 & 1981-2013 & 1.031 & 1.298 & 0.013 &                         \\
  UC  *  & 1949-1980 & 1.113 & 1981-2015 & 1.236 & 1.111 & 0.012 & long series             \\
  \hline
	\end{tabular}
\end{table}

We find that, out of the 28 stations, 24 indicate a higher ratio after 1980, while only one station indicates a constant ratio, and only three stations give a decreasing trend. As the curves have larger fluctuations, the timing of the transition is less accurate, but all series suggest a short transition within at most 3 years around 1980, preceded and followed by periods without systematic trends. 

The simple arithmetic mean of the listed SN ratios is 1.166 (range: 0.812 -- 1.731), which is larger than the jump amplitude calculated above. However, given the larger uncertainties in individual series and the different time intervals covered by each station, this mean value is only indicative. By keeping only stations with well-traced reliability, and which were active over most of the $F_{10.7}$ time interval (marked with a * in Table \ref{T:StationRatios}), we find a  mean value of 1.139 (range: 1.047\,--\,1.235), which is more reliable and comes reasonably close to the value found using the SN series as reference.

\emph{As this comparison involves only raw data from multiple independent observers, this verification thus allows us to confirm that the 1980 jump definitely occurs in the $F_{10.7}$ series and is not due to any unsuspected and uncorrected flaw in the SN series}.

Of course, another independent test could come from a comparison with equivalent radio flux measurements, made by another radiotelescope at the same wavelength or at a neighbouring one. \cite{NicoletBossy1985} made such a comparison with data from the Toyokawa station (Nagoya, Japan), but by then, the data extended only until 1982, thus only two years after the jump, which makes the detection of the jump difficult. More recently, another combined study by \cite{YayaEtal2017} also uses $F_{10.7}$ next to the Toyokawa series. Here also, there is no mention of a jump in 1980, but this study was focused on short-term predictions and did not look for such long-term inhomogeneities. So, a more focused study of long-term radio data series is still needed but goes beyond the scope of this article. 

\subsection{Possible cause of the 1980 jump: historical elements}

What could be the cause of this transition? \citet{TappingMorton2013} and \citet{Tapping2013} mention key dates in the history of the $F_{10.7cm}$ radio flux production. Regarding the receiver and facility, the greatest care was devoted to the calibration of the instrument, which was checked against the same horn-antenna references \citep{TappingCharrois1994}. The radio-telescope was re-located only two times, and this happened in 1962 (Ottawa to Algonquin Park) and 1990 (Algonquin Park to Penticton). None of those transfers left a detectable trace in the series, and in particular, those dates do not match the 1980 jump indicated by the data.
 
However, \citet{Tapping2013} mentions that in the late 1970’s, the original manual processing of the daily recordings was replaced by an automated procedure. This happened around the time when A.E.\,Covington, who had initiated the $F_{10.7}$ standard flux measurements and had directed its production continuously since 1947, went on retirement in 1979. This was thus the very first time a new team took over the task. This new team tried to automate the fully manual method that was applied until then to empirically eliminate the emission peaks caused by solar flares, but this computerized method was finally abandoned in 1985 because of insufficient reliability (Tapping 2019, private communication).

The 1979-1985 timing of this change of team and of post-processing method matches quite closely the moment when we find this scale jump. Based on similar disruptive events found in the history of the Z\"urich sunspot number, we suspect that this transition is the cause of the 1980 $F_{10.7}$ scale change. Indeed, only two highly dedicated scientists cared for the production of the $F_{10.7}$ index, each one for several decades: A.E.\,Covington 1947-1979 and K.\,Tapping 1985-present. Only once, in 1979, there was a transition when the experience and practices developed by Covington during the first 32 years had to be transmitted to the team that succeeded him. Part of the subtleties and habits may get lost in such knowledge transfers, especially as the processing procedure was largely manual until then, and proved to be difficult to convert into a computer algorithm. 

Moreover, the extraction of the $F_{10.7cm}$ background flux (the so-called ``S'' component) requires the elimination of flare-associated bursts, of radio-frequency interference, and of the sky background emission. Therefore, the final index depends on those post-processing steps, and not only on the proper calibration of the receiver and antenna. So, even if the raw fluxes were always accurately calibrated, this post-processing step may influence the final filtered index. Therefore, we can speculate that the changes in the methods that seem to have occurred just after 1979 could perfectly have caused this jump in the resulting index, without involving any technical flaw in the instrument itself and its calibration.
  
Moreover, our results indicate that there was no slow progressive drift of the $F_{10.7}$ flux relative to the sunspot number, but that a scale change occurred abruptly between two constant periods. This is in contradiction with the analysis by \citet{TappingMorgan2017}, who used a temporal curve fitting that did not allow to detect such a sharp transition.  This abrupt transition occurs at the maximum of cycle 21 and is unique over the last 6 solar cycles. Such a step-like jump does not match any known solar event in 1980 that would be unique over the last 70 years. Likewise, no mechanism generating the solar activity cycle can account for such an abrupt discontinuity, which require intrinsic timescales as short as one month. 

Therefore, we consider that it is very unlikely that the Sun itself induced this sudden change in the relation between the $F_{10.7cm}$ flux and the sunspot number, while a slow trend, as incorrectly diagnosed by \citet{TappingMorgan2017}, allowed such an hypothesis. By contrast, the production process and its historical evolution, retraced above, contain various elements that can induce such a sudden jump. So, this past history as recorded in archived documents deserve very careful attention, in order to validate or invalidate this processing issue.

\subsection{Solar-cycle and other modulations}
\begin{figure}
	\centering
	\subfigure{\includegraphics[width=1.\columnwidth,trim=0cm 0cm 0cm 0cm,clip]{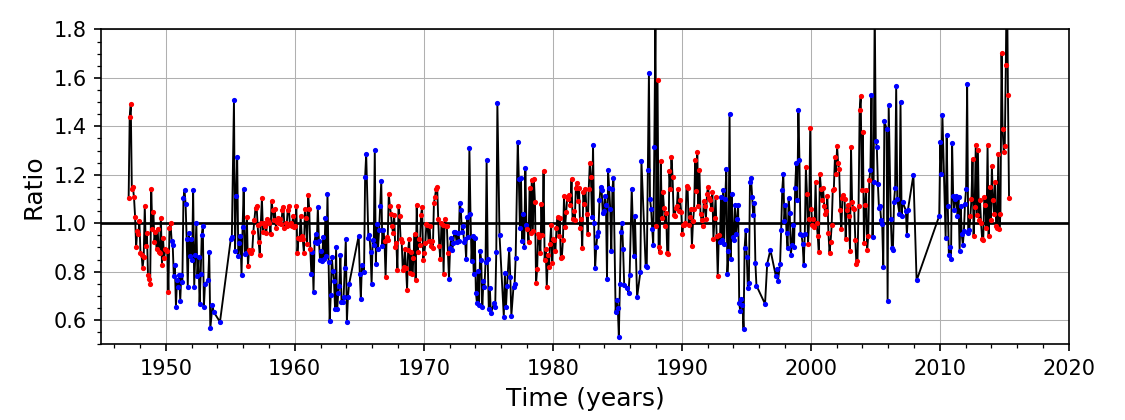}}
    \subfigure{\includegraphics[width=1.\columnwidth,trim=0cm 0cm 0cm 0cm,clip]{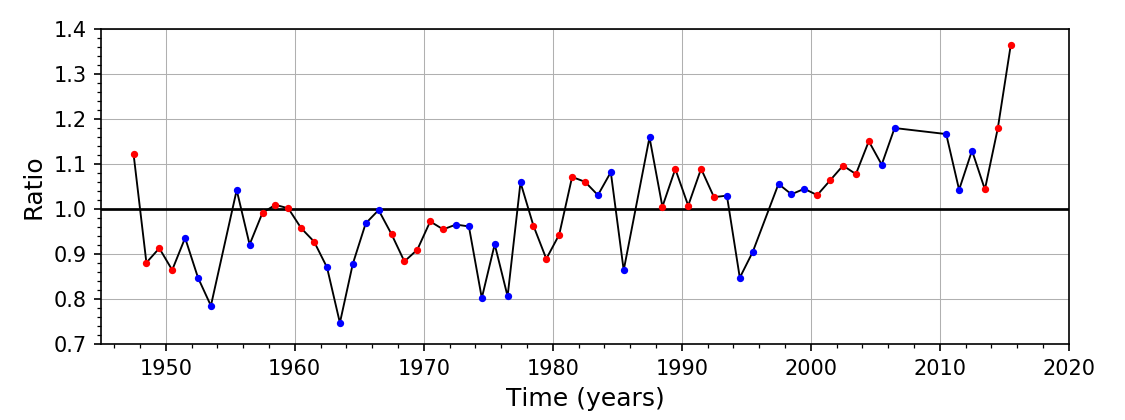}}
		\caption{\small 
			Temporal variation of the ratio between the monthly mean  $F_{10.7}$ flux and a constant proxy relation ($4^{th}$-order polynomial) (upper panel), after subtracting a 67 sfu quiet-Sun base flux. The same ratio for the yearly mean $F_{10.7}$ flux is shown in the lower panel. The points near cycle maxima and minima are colored in blue and red respectively. The monthly means show a slight solar-cycle modulation, and both curves show the 1980 jump between two stable periods.		
			\label{F:RatioTime}} 
\end{figure}

As noted for in Section \ref{S:MeanProf}, the mean daily $F_{10.7}$ flux for a given $S_{\mathrm{N}}$ shows a significant difference between the minima and maxima of the solar cycle. This cycle modulation is largest for raw daily data (Figure \ref{F:Hist_d}) and decreases for longer time scales. In order to check if this modulation is affecting the monthly and yearly means used to build the long-term proxy, we computed the ratio between monthly or yearly mean values and a single constant fit: the $4^{th}$-order polynomial fitted to the entire series, given in Tables \ref{T:PolyOLSm} and \ref{T:PolyOLSy}. Any deviation from a constant relation will appear as a deviation from unity. The result is shown in figure \ref{F:RatioTime}, where we marked the periods around maxima and minima of the solar cycles by blue and red dots.

The plot for the monthly means (Figure \ref{F:RatioTime}, upper panel) is dominated by random month-to-month variations. No annual modulation is present, indicating that the adjusted $F_{10.7}$ was accurately converted to the 1AU distance. The 1980 jump between two stable intervals is also clearly visible. Now considering the solar cycle, a slight modulation can be found: although the ranges of monthly values around maxima and minima largely overlap, the range near maxima is slightly higher than around minima. The difference is subtle and remains smaller than the other deviations mentioned above.

Finally, the ratios for yearly means do not show any cycle modulation (Figure \ref{F:RatioTime}, lower panel). The only clear systematic variation is the 1980 jump. Virtually all points are below 1 before 1980 and above 1 after 1980, confirming again clearly the jump and the absence of any progressive trend.

Overall, we can thus confirm that the solar-cycle modulation is playing a significant role only for timescales shorter than a month. We can interpret this effect by invoking the same mechanisms as the ones explaining the change of average background during spotless periods (cf. Section \ref{S:BackgndFlux}). As activity increases, the plage component can contribute to a flux excess when sunspot activity drops momentarily, because the facular and plages associated to all active regions persist for a much longer time than the corresponding active regions. The higher activity prevailing around those dips in the sunspot number thus prevents the $F_{10.7}$ to decrease as sharply. Therefore, the net effect must always be an excess, which matches the upper tail of the daily distribution in Figure \ref{F:Hist_d}. Near the minima of cycles, given the small number of active regions, this persisting background is largely absent, thus giving a smaller plage excess, leading to the observed solar-cycle variation in Figure \ref{F:Hist_d}. As this temporal smearing effect corresponds to the lifetime of plages, which ranges from weeks to a few months, it should vanish for long timescales, like we find in our analysis (Figure \ref{F:RatioTime}).

\section{Conclusions}
Summarizing, our analysis brings the following conclusions regarding the global $F_{10.7}/S_{\mathrm{N}}$ proxy relation:
\begin{itemize}
	\item No previously published $F_{10.7}$ proxy relation is fully satisfactory. Existing proxies deviate from the data points either in the low or high range, though there is a fair agreement in the intermediate linear range. Those proxies are also lacking error bars, limiting their applicability.
	
	\item The $F_{10.7}/S_{\mathrm{N}}$ relation is fully linear within uncertainties from the lowest to the highest observed values, when taking raw daily values without any temporal averaging. The $F_{10.7}$ flux only deviates from the linear relation for $S_{\mathrm{N}}$ below 11 (single spot) and even $S_{\mathrm{N}}=6$. $F_{10.7}$ reaches a lower base background only when the Sun is spotless.
	
	\item When working with monthly and yearly means, the $F_{10.7}/S_{\mathrm{N}}$ relation becomes non-linear in the low range, for $S_{\mathrm{N}}$ below 30 to 50. This non-linearity can be fully explained by the effect of temporal averaging on daily data consisting of a fully linear relation, plus a lower background (fully inactive Sun). A $F_{10.7}$ proxy relation is thus only valid for a specific temporal averaging of the base daily data.
	
    \item  A $4^{th}$ degree polynomial gives the best fit to the monthly mean data, in particular the non-linear section below $S_{\mathrm{N}}=50$ down to 0. A linear function is sufficient for all $S_{\mathrm{N}}$ values above about 30. 
    
    \item We derived standard errors $\sigma_p$ on the polynomial values. As a direct mathematical error-propagation calculation does not exist taking into account the inter-dependencies between the least-square polynomial coefficient, those errors were derived empirically from the data by determining the conditional errors for each separate term of different degree in the polynomial. For practical applications, we also assembled a simple mathematical formula that closely approximates those data-based conditional errors.
\end{itemize}
    
In addition, we derived new properties of the $F_{10.7}$ quiet-Sun background flux:    
\begin{itemize}
	\item This background flux for a spotless Sun depends on the duration of the spotless interval. Its mean value is 68 sfu for long inactive periods, but rises to 74 sfu for a spotless duration of one day. With only a few isolated exceptions, 67 sfu is the lowest flux value, independently from the spotless duration, but the mean quiet-Sun background is always higher.

	\item Given the actual duration of spotless periods, a temporal averaging over one month is close to optimal to reflect the lowest range of $F_{10.7}$ background flux, with 68 or 69 sfu as the lowest mean background over one month.
    \item A $F_{10.7}$ excess flux is present in raw daily data and induces a 12\% solar-cycle modulation. The latter vanishes at long timescales, and is already barely detectable in monthly averages ($\approx 3\%$). This implies that a single standard $F_{10.7}$ -- $S_{\mathrm{N}}$ proxy relation, independent of the phase of the solar cycle, can only be derived from temporal scales longer than about one month, and will only be fully accurate for those timescales.
\end{itemize}

Finally, by checking for any temporal variability of the $F_{10.7}/S_{\mathrm{N}}$ relation over the entire duration of the data series, we found a significant inhomogeneity:
\begin{itemize}
	\item The $F_{10.7}$ series is affected by an upward jump in 1980, separating two stable periods without other jumps or trends.  Relative to the SN series, the $F_{10.7}$ index is 10.5\% higher after 1980. Monthly values suggest that the jump occurred at the transition between two calendar years, i.e. between December 1980 and January 1981.
	
	\item In order to exclude a possible flaw on the side of the sunspot number series, which also went through a methodological transition in 1980--1981, we compared the $F_{10.7}$ with raw Wolf numbers from a large sets of independent stations. With only a few exceptions, most of them indicate that $F_{10.7}$ becomes higher after 1980. This thus firmly establishes that the scale jump belongs to the $F_{10.7}$ time series.
	
	\item The jump is abrupt and makes any interpretation in terms of a true solar effect difficult. On the other hand, this abrupt transition happens close to important changes in the observing team (retirement of A.E. Covington), and when changes were introduced in the post-processing method (computerization of an originally manual processing). Just like the diagnosed jump, this operational transition is unique in the history of the $F_{10.7}$ production.
	\end{itemize}

This study thus indicates that we can still learn a lot about this fundamental long-term solar-activity index. The consistent relation between the $F_{10.7}$ quiet-Sun background and the duration of the quiet period, the role of temporal averaging on the non-linearity of the proxy relation as well as the cycle-dependent excess flux found only in daily data, all invite us to pay more attention to temporal scales. Our results confirm the mixed contribution of two components in $F_{10.7}$, one from sunspots and another one from weaker magnetic fields primarily in plages. The latter introduces a time-diluted variation relative to the initial magnetic flux emergence in active regions, recorded by the SN. As this source-mixing in $F_{10.7}$ plays a role only at short time-scales, from days to weeks, it goes beyond the scope of this long-term proxy study. Still, this aspect calls for more attention in futures analyses of $F_{10.7}$, and it should also be kept in mind in all uses of $F_{10.7}$ for space-weather or space-climate applications.
 
Given the inhomogeneity in the $F_{10.7}$ time series that we diagnose here in detail, it is clear that our new global polynomial proxy does not reproduce optimally the actual relation before and after 1980. Until a correction is adopted for the $F_{10.7}$ data, it must be considered as the best overall sunspot-based proxy. The slope and curvature are largely valid for the whole series, but a different scale factor must be applied for each half of the series: proxy values will be about 5\% too high before 1980 and 5\% too low for the more recent years relative to the current version of the $F_{10.7}$ series. The polynomials derived for each half of the series in Table \ref{T:TwoPoly} can be used in applications focusing only on time periods before or after 1980.

The combined historical evidence indicates that a thorough analysis of the production process of the $F_{10.7}$, in particular in the late 1970's to early 1980's, is needed to clarify any possible change, and potentially to find a self-consistent correction to restore the homogeneity over the entire series since its beginning. If the archived $F_{10.7}$ data prove to be insufficient, our best second option would be to use the long overlap with the sunspot number series. The latter may provide an even better reference in the future, as a new re-calibration is in preparation, which could improve in particular the $S_{\mathrm{N}}$ values from the Z\"urich period before 1981. 

\begin{acknowledgements}
      This work and the team of the World Data Center SILSO, which produces and distributes the international sunspot number used in this study, are supported by Belgian Solar-Terrestrial Center of Excellence (STCE) funded by the Belgian Science Policy Office (BelSPo). This work also partly benefited from the joint work of the International Team 417 ``Recalibration of the Sunspot Number Series'', funded by the International Space Science Institute (ISSI, Bern, Switzerland) and chaired by M. Owens and F. Clette. We also wish to acknowledge the particularly useful, informative and friendly discussions with Ken Tapping, who is the dedicated guardian and the living memory of the $F_{10.7cm}$ radio flux.
\end{acknowledgements}



\end{document}